\documentclass[manuscript]{emulateapj}
\usepackage{graphicx}
\usepackage{amsmath}
\usepackage{amssymb}
\usepackage{wasysym}
\usepackage{color}

\shorttitle{The Lives of Stars}
\shortauthors{Vickers \& Smith}

\begin{document}
\author{John J. Vickers\altaffilmark{1,$\dagger$}, Martin C. Smith\altaffilmark{1}}

\title{The Lives of Stars: Insights From the TGAS-RAVE-LAMOST Dataset}

\altaffiltext{$\dagger$}{johnjvickers@shao.ac.cn}

\altaffiltext{1}{Key Laboratory for Research in Galaxies and Cosmology, Shanghai Astronomical Observatory, Chinese Academy of Sciences, 80 Nandan Road, Shanghai 200030, China}

\def\mean#1{\left< #1 \right>}

\begin{abstract}

In this paper we investigate how the chemical and kinematic properties of stars vary as a function of age. Using data from a variety of photometric, astrometric and spectroscopic surveys, we calculate the ages, phase space information and orbits for $\sim$125,000 stars covering a wide range of stellar parameters.

We find indications that the inner regions of the disk reached high levels of enrichment early, while the outer regions were more substantially enriched in intermediate and recent epochs. We consider these enrichment histories through comparison of the ages of stars, their metallicities, and kinematic properties, such as their angular momentum in the solar neighborhood (which is a proxy for orbital radius). We calculate rates at which the velocity dispersions evolve, investigate the Oort constants for different aged populations (finding a slightly negative $\partial V_{C} / \partial R$ and $\partial V_{R} / \partial R$ for all ages, being most negative for the oldest stars), as well as examine the behavior of the velocity vertex deviation angle as a function of age (which we find to fall from $\sim$15 degrees for the 2 Gyr aged population to  $\sim$6 degrees at around 6.5 Gyr of age, after which it remains unchanged). We find evidence for stellar churning, and find that the churned stars have a slightly younger age distribution than the rest of the data.

\end{abstract}

\section{Introduction}\label{sec:introduction}

The study of the history of the Milky Way Galaxy, galactic archaeology, relies on fossil evidence to draw conclusions about the formation processes that resulted in the Milky Way we see today. Some of this evidence can be seen directly in stellar number counts, as is the case for stellar streams formed both by the infall of large galaxies (e.g. the Sagittarius dwarf spheroidal; \citealt{iba1994}, \citealt{her2017}) and small systems (e.g. the tidal tails around the globular cluster Palomar-5; \citealt{ode2001}, \citealt{erk2017}). Such evidence provides direct observational constraints on the extent to which the Milky Way is formed in situ \citep{egg1962}, and which portion is formed via accretion (\citealt{sea1978}, \citealt{hel2017}).

Evidence of formation mechanisms can be seen indirectly in the form of the Galactic warp (\citealt{djo1989}, \citealt{sch2017ii}) and flare (\citealt{ken1991}, \citealt{fea2014}), which could be caused either by accretion or by internal disk instabilities. Internal disk instabilities could also give rise to features such as the Galactic bar (\citealt{wei1992}, \citealt{com1981}), and the spiral arms (which are perhaps long lived, or perhaps transient; for discussion, see, for example, \citealt{sel2011} or \citealt{mar2017}).

In the era of photometric surveys, it was difficult to draw sweeping conclusions about these broad features, but the more one looked, the more evidence pointed to a fairly chaotic formation (see for example the lumpy Milky Way topography in \citealt{new2002}, or the complex Magellanic system in \citealt{bel2016}). Estimates of kinematics could be made based on 2-dimensional astrometric proper motions; however, until recently, these have been defined by long-baseline observations cross-matching the most modern photometric observations with 20th century photographic plate surveys -- see for example the UCAC \citep{zac2013} and PPMXL \citep{roe2010} catalogs -- which generally achieve precisions of $\sim$3 mas yr$^{-1}$, or 15 km s$^{-1}$ kpc$^{-1}$ (an inconvenient limit for discussing features several kpc away). Other problems also occur in the confusion of crossmatching large numbers of surveys with varying specifications, leading to some discrepancies between proper motion catalogs \citep{vic2015}, systematics within individual catalogs (\citealt{wu2011}, \citealt{vic2016}), and, for some, errors clustered around the plate-scale indicating the sub-optimal condition of crossmatching plates with CCD data \citep{pea2017}.

In the early 2000s, a variety of large scale spectroscopic surveys began producing data:
\begin{itemize}
\item the Radial Velocity Experiment released almost 25 thousand spectra in 2006 (RAVE; currently $\sim$520 thousand spectra available; \citealt{ste2006}, \citealt{kun2017}),
\item the Sloan Digital Sky Survey (SDSS; \citealt{yor2000}) project the `Sloan Extension for Galactic Understanding and Exploration' (SEGUE; \citealt{yan2009}) released spectra of 240 thousand stars in 2009 (SEGUE-2 added another $\sim$120 thousand),
\item the Large Area Multi Object Spectroscopic Telescope (LAMOST; \citealt{luo2015}) released 1.5 million spectra in 2013 (currently $\sim$7.5 million spectra are available),
\item The Apache Point Observatory Galactic Evolution Experiment (APOGEE; \citealt{hol2015}) released spectra of 150 thousand stars in 2015, and APOGEE-2 is currently surveying a further 300 thousand.
\end{itemize}
With this outpouring of data, the parameter spaces of radial velocity and abundance became available for large numbers of stars. With radial velocities, overdensities could be tested for coherence at distances impossible with extant proper motions (a nice example of this is seen in the 34 kpc distant Cetus Polar stream of \citealt{new2009}), and the potential of the Milky Way could be tested out to tens of kpc (as done with blue horizontal branch stars by \citealt{xue2008}).

With abundances, work could be done by comparing elemental abundance ratios to provide  indirect insights into the ages of populations. The ratios of iron and alpha-element abundances particularly holds information about star formation rates of populations (see, for example \citealt{whe1989} and references therein) and is useful for exploring differences between spatially mixed populations, such as those of the canonical thin and thick disks \citep{gil1983}, which are thought to have different formation histories and thus chemistry (see \citealt{fuh1998}, and later \citealt{bov2012} and \citealt{bov2013}). Alpha elements, being primarily produced in swift, type-II supernovae, are generally thought to be indicative of high star formation regions. While iron is injected into the interstellar medium by type-Ia supernovae which occur in all disk regions at a fairly constant rate. So stars with high alpha abundances relative to iron, such as thick disk stars, were probably born in a quickly evolving, high star formation environment; and the alpha abundance relative to iron gradually falls for all subsequent generations of stars born into a more quiescent Milky Way being slowly enriched in iron by type-Ia supernovae. More complicated ``chemical tagging" (see \citealt{fre2002}),  utilizing a wider variety of elemental abundances arising from different nuclear processes, can provide even more information, and is explored early in F and G disk stars in \citet{ben2003} and \citet{ben2005}, or more recently in, e.g., \citet{tin2016} in APOGEE or \citet{kos2018} in GALAH (Galactic Archaeology with HERMES, the High Efficiency and Resolution Multi-Element Spectrograph; \citealt{des2015}). Recently, chemodynamical models of Milky Way like galaxies have been crucial to understanding these abundance observations in the context of cosmological galaxy evolution with effects such as inside-out formation scenarios (\citealt{pil2012}), migration (\citealt{kub2015i}, \citealt{kub2015ii}, \citealt{min2014}), spiral arm influence (\citealt{gra2016}), and realistic galactic environments (\citealt{ma2016}).

However, while many proxies for age existed -- such as color (with extremely blue colors generally singling out stars with very short lifespans), color-apparent-magnitude (with coherent populations following isochrones in this space), and alpha-iron ratios (making certain assumptions about the formation mechanisms leading to those ratios) -- precision ages via isochrone matching remained out of reach for most stars until accurate distances could be obtained. Age estimates may be done with atmospheric parameters alone, but the precision is generally low in large spectroscopic surveys.

Some other specialized methods for measuring age exist, such as elemental abundance ratios in giants \citep{mar2016}, or astroseismic age-estimation for giants \citep{cas2016}, but robust isochrone matched age-determination for the full range of stellar types had to wait until more accurate distance information became available.

With the Hipparcos mission (\citealt{per1997}, \citealt{van2007}) providing parallaxes for 2.5 million stars in the form of the Tycho-2 astrometric catalog \citep{hog2000}, the way was open to directly calculate the ages and kinematics for massive numbers of stars, as was done by the Geneva-Copenhagen Survey \citep{nor2004} for about 14,000 stars. Similar endeavors were undertaken using the High-Accuracy Radial velocity Planetary Searcher \citep{may2003} for about 1,000 stars (see for example \citealt{hay2013}).

The Gaia space telescope \citep{gaia}, launched in 2013, will drastically improve the current astrometric precision on parallaxes (with micro-arcsecond parallaxes expected out to 20$^{th}$ magnitude) and proper motions (estimated to be two to three orders of magnitude more precise than current large, ground-based surveys). This will greatly increase the sample of stars for which 6-dimensional phase space information, and ages, can be accurately calculated by crossmatching with existing spectroscopic surveys. The mission will also provide $\sim$150 million radial velocities, atmospheric parameters for $\sim$5 million stars, and detailed abundances for $\sim$2 million stars (according to the science performance page\footnote{https://www.cosmos.esa.int/web/gaia/science-performance}; see also \citealt{rec2016}).

This paper presents an analysis of the kinematic and abundance trends as a function of age for a large sample of stars in a kpc wide sphere around the Sun, taking advantage of the first Gaia data release, the fourth LAMOST data release, and the fifth RAVE data release.

We will consider the behavior of stars in different age groups with a specific interest in stellar migration. Stellar migration may generally be thought of as consisting of two mechanisms: 1) the gradual heating of stars off of circular orbits and onto eccentric orbits, extending the radial range they inhabit but conserving angular momentum (so they travel quickly at perigalacticon, and slowly at apogalacticon), this causes populations to be observed over larger ranges of radii over time, an effect called ``blurring"; and 2) the exchange of angular momentum between individual stars and non-axisymmetric elements of the potential, such as the spiral arms or bar, allowing stars to change their orbital radii while maintaining near-circular orbits, an effect known as ``churning". Both of these mechanisms of stellar migration tend to flatten metallicity gradients. This nomenclature is taken from \citet{sch2009}.

Stellar mixing, or the net movement of stars away from their birth radii, has been widely investigated in simulations in recent years. Mixing was investigated as an effect of angular momentum exchange over spiral arm corotation radii (by e.g. \citealt{sel2002}, \citealt{ros2012}, \citealt{gra2014}) and was found to be responsible for large changes in individual stars' angular momenta, sometimes as large as 50\%. \citet{ros2008} looked into the effects of this process on observed age and abundance and found that this type of migration typically flattened the age-metallicity relationship. \citet{min2010} expanded on this with the addition of a bar component and found that bars may affect angular momentum change in the vicinity of the corotation radius and the outer Lindblad resonance and also that bars may enhance the rate at which metallicity gradients flatten \citep{min2011}. Further, \citet{ver2014}, using a disk with spiral arms embedded in a dark matter halo, showed that churning was most efficient on stars with low vertical velocity dispersions.

In Section \ref{sec:data} we outline the data used and describe corrections and cuts needed to clean it. In Section \ref{sec:age_estimation} we describe the process for estimating the ages and age errors for the stars. Section \ref{sec:coords} details the coordinates and orbit calculations. In Section \ref{sec:dynamics} we focus on the dynamics, discussing: the vertex deviation angle, the rate of increase of velocity dispersion and velocity dispersion ratios, the Oort constants and rotation curve as a function of stellar age, and some strange behavior of the youngest stars in our sample. Section \ref{sec:abundances} folds in abundance information to investigate the chemical evolution, looking at: inside out formation, the relative rates of enrichment for the inner and outer disk, blurring, and possible observational evidence for outward stellar churning. In Section \ref{sec:conclusions} we conclude.  An appendix is devoted to dealing with the selection functions of spectroscopic surveys; and another details some offsets and corrections between the LAMOST and RAVE data sets.

\begin{figure*}
\includegraphics[width=\linewidth]{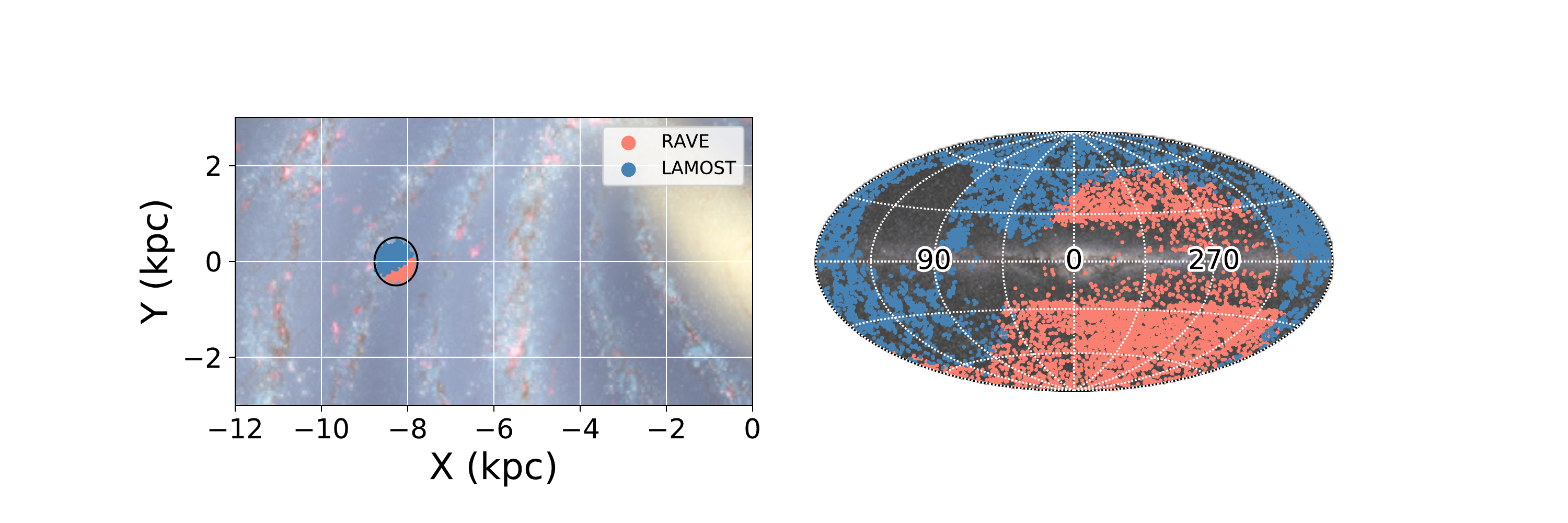}
\caption{The observational footprint of our combined dataset in relation to the Galaxy (\emph{left}) and on the sky (\emph{right}). A subset of 500 stars each from RAVE and LAMOST is used in the left panel, and a subset of 5000 each is used in the right panel to reduce the figure size. The black circle on the left is centered on the solar position with a radius of 500 pc.\\
The background images are taken without endorsement from: \emph{left}) NASA/JPL-Caltech/ESO/R. Hurt\footnote{http://www.eso.org/public/images/eso1339g/}, and \emph{right}) ESO/S. Brunier\footnote{http://www.eso.org/public/images/eso0932a/}.
}
\label{fig:footprint}
\end{figure*}

\section{Data}\label{sec:data}

The astrometric data used are the Tycho-Gaia Astrometric Solution \citep{tgas}. This dataset is a combination of the Tycho-2 astrometric catalog and the astrometric data collected in the first year of the Gaia satellite's operations. From these data we make use of the parallaxes ($\sim$0.3 mas uncertainties), the proper motions ($\sim$1 mas yr$^{-1}$ uncertainties), and the mean $G$ band magnitude ($\lesssim$0.03 mag uncertainties for our objects).

It has been noted by many that the transformation from parallax space into distance space $d$ is nonlinear and therefore can dramatically skew the error. \citet{bai2015} show that this effect will become pronounced after fractional parallax errors of 20\%. \citet{sch2017} have noted the distance estimates of TGAS are relatively consistent with the predictions of their internally consistent galactic model up to a fractional parallax error of 20\%. To correct for this it is necessary to adopt a prior based on estimated density distributions, as in \citet{ast2016}, or on density and velocity distributions as in \citet{sch2017}. We prefer to work here without a prior since we will be limiting ourselves to high precision data anyways, to get the most accurate age estimations. In light of this, we adopt 20\% parallax errors as a hard cut. In the future however, as \citet{bai2015} note, roughly 80\% of the Gaia catalog will be beyond this 20\% error threshold, so it is worth considering, now, the types of priors which will be acceptable to use to correct for this.

The spectroscopic data used come from two sources: LAMOST DR4v1, and RAVE DR5.

The LAMOST spectrographs collect R$\sim$1800 spectra in the wavelength range from 370-900 nm. As of DR4, the LAMOST data set contains over 4 million spectra above $\delta\sim-10^{\circ}$, with a particular spatial concentration in the direction of the Galactic anticenter. Of these, 70\% have signal to noise ratios above 20 in the $g$ band. The median errors are: $\delta$ T$_{eff} \sim$139.5 K; $\delta$ log($g$) $\sim$0.46 dex;  $\delta$ [Fe/H] $\sim$0.17 dex.

We crossmatch these LAMOST spectroscopic data with the TGAS data within 10" and we find that about 75\% and 97\% of the matches are within 1" and 4", respectively. We compare the (B-V) Tycho-2 colors with the LAMOST effective temperatures and find that the crossmatches out to 4" are not of a worse overall qualitative match than the $<$1" sample, so we choose 4" as the matching radius.

We further discard all objects with failed error estimates in T$_{eff}$, log($g$), or [Fe/H] (meaning pipeline failures, we require errors $>$ 0), and objects with signal to noise ratios in $g$, $r$, or $i$ less than 20. A final cut requiring positive parallaxes leaves us with $\sim$73,000 objects whose median errors are: $\delta$ T$_{eff} \sim$83.6 K; $\delta$ log($g$) $\sim$0.40 dex;  $\delta$ [Fe/H] $\sim$0.10 dex. However it should be noted that the pipeline errors from the LAMOST data products are known to be flawed (see for example \citealt{sch2017}). We change the radial velocity errors to be in line with expected values from comparisons of the LAMOST kinematics to models (7.2 km s$^{-1}$, \citealt{sch2017}). We leave the gravity, temperature, and [Fe/H] error values at their pipeline values, which are likely overestimated (based on comparison with RAVE measurements and error distributions for a population with observations existing in both surveys).

The RAVE spectrograph observes at a higher resolution (R$\sim$7500) in a smaller wavelength range (841-879.5 nm) around the calcium triplet lines. These observations are arranged relatively evenly in pencil beams south of $\delta\sim0^{\circ}$, excepting in the plane of the Galaxy. The RAVE data product boasts more precise pipeline error medians than LAMOST -- $\delta$ T$_{eff} \sim$74 K; $\delta$ log($g$) $\sim$0.16 dex;  $\delta$ [M/H] $\sim$0.10 dex.

Rather than crossmatching the TGAS data to the RAVE data ourselves, we utilize the helpfully included crossmatch which is packaged with RAVE DR5. We place similar restrictions on the data set: a positive parallax, total parallax error $<$20\%, signal to noise ratio $>$20, non-zero errors on T$_{eff}$, log($g$), and [M/H] measurements. We additionally require the flags `c1,' `c2,' and `c3' to be `n' for normal. This yields a sample of about 67,000 objects with atmospheric parameter errors around: $\delta$ T$_{eff} \sim$76 K; $\delta$ log($g$) $\sim$0.14 dex;  $\delta$ [M/H] $\sim$0.10 dex.

By using data from these two surveys, we essentially double our spectroscopic sample which overlaps with the TGAS data,  and also gain almost-full-sky coverage rather than the half-sky coverage offered by either survey independently (see Figure \ref{fig:footprint}).  However it should be noted that this creates an extremely complicated selection function, with RAVE generally observing brighter, cooler objects more oriented toward the center of the Galaxy than LAMOST. Addition selection complexity is added with respect to the Gaia to Tycho-2 matching, and the subsequent matching of that catalog to LAMOST and RAVE. We have endeavored to address this in as comprehensive a way as possible in Appendix \ref{app:sel_fn}.

Duplicate observations exist in the data; they are caused by overlaps between the RAVE and LAMOST targets, and also by intentionally repeated observations. We remove these duplicates from our sample by selecting only the objects with the lowest atmospheric temperature, gravity, and metallicity errors (preferentially in that order) within a 7" circle on the sky.

We note that the RAVE [M/H] measurement is not wholly interchangeable with the LAMOST [Fe/H] estimate and also that the radial velocity estimates are systemically offset between the two surveys. These issues are addressed in Appendix \ref{app:scale}.

\section{Age Estimation}\label{sec:age_estimation}

We estimate the ages of our stars using a Bayesian method similar to that outlined in \citet{jor2005}.

An isochrone grid is constructed using the Padova isochrones\footnote{http://stev.oapd.inaf.it/cgi-bin/cmd} (see \citealt{bre2012}, \citealt{che2014}, \citealt{che2015}, \citealt{tan2014}). The grid is evenly spaced in [M/H] every tenth of a dex from $-1.5$ to $0.3$ dex. The grid is non-uniformly spaced in age: every $0.01$ Gyr from $0.03$ to $0.1$ Gyr; every $0.05$ Gyr from $0.1$ Gyr to $0.5$ Gyr; every $0.1$ Gyr from $0.5$ to $1$ Gyr; every $0.25$ Gyr from $1$ to $4$ Gyr; every $0.5$ Gyr from $4$ to $13$ Gyr. We space the grid in this uneven manner because isochrone morphology changes with age less at older ages than at younger ages.

For each star in our sample, we use the parallax, parallax error, and observational $g$-band Gaia magnitude ($G$) from the TGAS dataset. If the object is in the RAVE dataset, we use the stellar parameter pipeline values of the [M/H], log(g) and T$_{eff}$; if the object is in LAMOST, we use the estimated [M/H] from Appendix \ref{app:scale} and the pipeline values of log(g) and T$_{eff}$.

The isochrones are converted into parallax space based on the individual star for which the age is currently being estimated by considering the observed $G$ magnitude of the star and the absolute $G$ magnitude of the isochrone:

\begin{equation}\label{eqn:plx}
\pi_{iso} = 100 \cdot 10^\frac{G_{abs} - G_{obs}}{5}.
\end{equation}

In the paper, when considering the errors on the parallaxes, we do \emph{not}, as is sometimes seen, add in a 0.3 mas systematic error. This is following the recommendation of \citet{bro2017}.

Each isochrone consists of $n$ points, such that $i = n-1$ sequential line segments may be constructed. $\chi^{2}$ distances to each of these $i$ line segments are calculated:

\begin{equation}\label{eqn:chisq}
\chi^{2}_{i} =  \sum_{j}{\left(\frac{q_{j}^{obs} - q_{i,j}^{iso}}{\sigma_{j}^{obs}}\right)^{2}},
\end{equation}

where $j$ indicates the parameters ($\pi$, [M/H], log(g) and T$_{eff}$); $q^{obs}$ and $\sigma^{obs}$ are the measurement and measurement errors of the parameter; and $q^{iso}_{i}$ corresponds to the parameter values of the closest point on the $i^{th}$ line segment connecting points $i$ and $i+1$ in the isochrone. The $\chi^{2}$ values are then converted to likelihoods as:

\begin{equation}\label{eqn:likelihood}
L_{i} = \prod_{j}{ \frac{ 1 }{ \sqrt{2\pi}\sigma_{j} } } \times exp\left( -\frac{ \chi_{i}^{2} }{ 2 } \right).
\end{equation}

We further weight the likelihoods by an initial mass function of the line segment $i$ of the isochrone so:

\begin{equation}\label{eqn:likelihood_weighted_integral}
L_{iso} = \int_{M} L_{i} \cdot IMF(M) \cdot dM,
\end{equation}

such that

\begin{equation}\label{eqn:lk}
L _{iso} = \sum_{i} L^{'}_{i},
\end{equation}

where

\begin{equation}\label{eqn:likelihood_weighted_sum}
L^{'}_{i} = L_{i} \cdot IMF\left(\frac{M_{i} + M_{i+1}} { 2 }\right) \cdot (M_{i+1} - M_{i}).
\end{equation}

Here, $M_{i+1}$ and $M_{i}$ are the \emph{initial} masses of the isochrone points bounding isochrone line segment $i$ and

\begin{equation}\label{eqn:kroupa1}
IMF(M) \propto M^{-\alpha}
\end{equation}

is the \citet{kro2001} initial mass function. $\alpha$ in this case is:

\begin{equation}\label{eqn:kroupa2}
   \alpha(M) = \left\{
     \begin{array}{ll}
       0.3 & : M < 0.08M_{\odot} \\
       1.3 & : 0.08M_{\odot} < M < 0.5M_{\odot}\\
       2.3 & : M > 0.5M_{\odot}.
     \end{array}
   \right.
\end{equation}

Marginalizing over the metallicity, we find the posterior probability of a star being age $\tau$ to be:

\begin{equation}\label{eqn:posterior_integral}
G(\tau) = \int_{All [M/H]} L_{iso}(\tau, [M/H])\cdot d[M/H].
\end{equation}

The 68\% confidence intervals on the ages are estimated, as suggested in \citet{jor2005}, by finding the extent of the $G(\tau)$ function greater than 0.61 times the maximum value of $G(\tau)$.

\begin{figure}
\includegraphics[width=\linewidth]{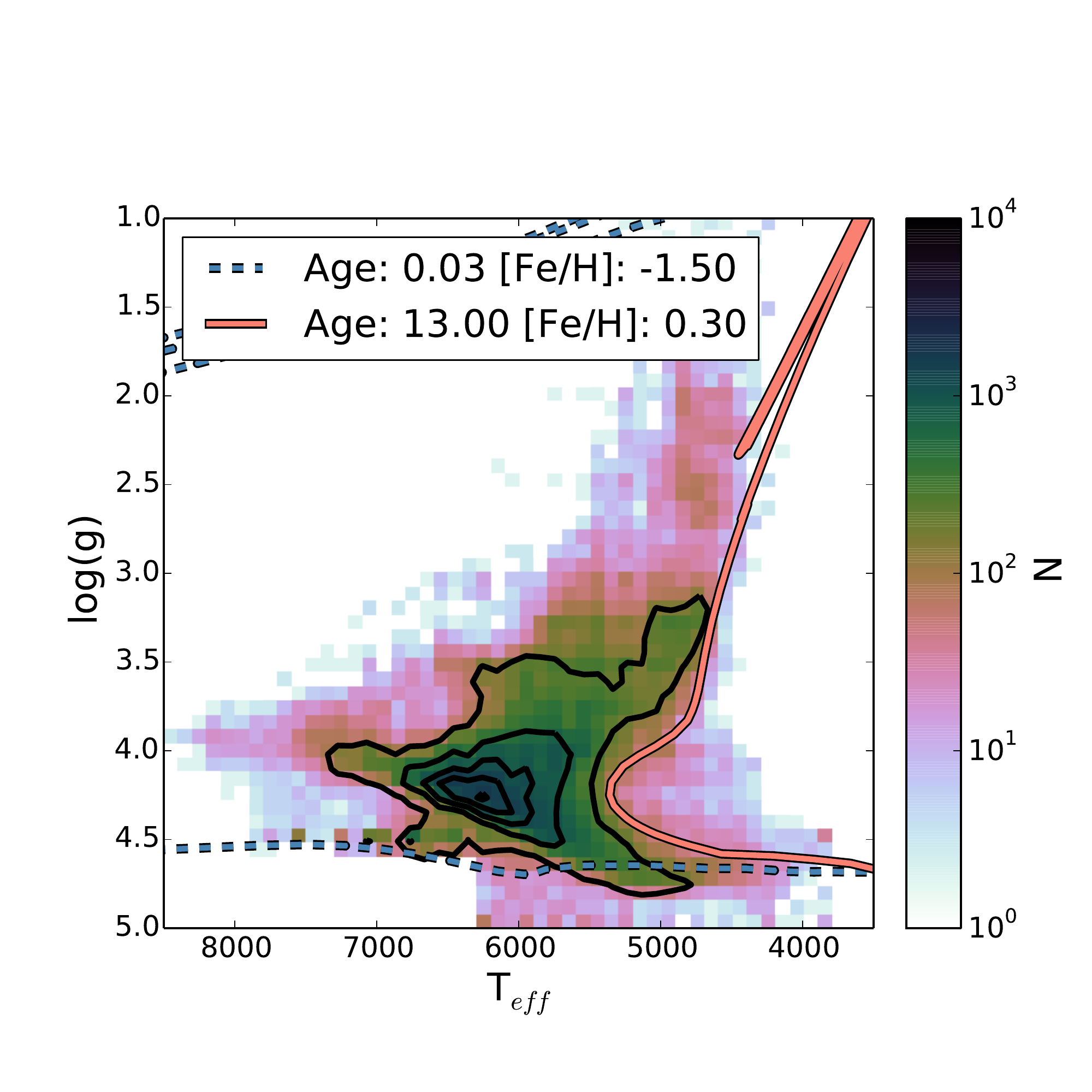}
\caption{After calculating the relative ages of the stars in our dataset, we found excesses in the edge-case age bins of 0.03 Gyr and 13 Gyr. This is probably owing to observational errors scattering the stars out of the range of our hottest (dashed blue line) and coolest (solid red line) isochrones. The contours are at 5\%, 25\%, 50\%, 75\% and 95\% of the maximum bin value.}
\label{fig:outlier_cmd}
\end{figure}

At this point, we implement some additional data quality cuts. In our initial age distribution, we have excesses at the edge cases of our data, the reason for this is apparent from Figure \ref{fig:outlier_cmd}. This Hess diagram depicts the observational sample, along with the ``reddest" and ``bluest" isochrones in our grid. It is apparent that a non-negligable portion of the observed stars lie outside the search space, for example in the crook redward of the old isochrone turnoff and the high log(g) observations below the main sequence. This is not especially surprising, considering the surface gravity errors of $\sigma$log(g) $\sim$ 0.3 dex and temperature errors of $\sigma$ T$_{eff}$ $\sim$ 100 K would naturally scatter the observations off of their true locus. The stars outside of this search space will be placed into the edge-case bins, overpopulating them. In light of this, we further reject any star which is 3$\sigma$ inconsistent with the nearest point on the isochrone grid. A sample of about 105,000 objects remains. Rejecting the youngest (and oldest) edge-case age bins leaves $\sim$98,000 ($\sim$87,000) stars.

Further, we find that the weighting scheme implemented in Appendix \ref{app:sel_fn} produces weights which follow a near-normal distribution in log-space. The weights vary by a few orders of magnitude and we wish to prevent highly weighted stars from completely overpowering the rest of the sample. Based on these two observations, we cut the data such that the allowed weighting is within three standard deviations of the mean of the calculated weight values (in logarithmic space). This leaves $\sim$55,000 stars.

We briefly explore the effects of extinction on our sample by applying the full column density extinctions of \citet{sch1998}, with the extinction coefficient of $A_G = 2.55\cdot E(B-V)$ from \citet{bel2017} to our data and recalculating the ages. We find that $\tau_{observed}-\tau_{dereddened} = 0.08 \pm 1.89 Gyr$. Since the systematic difference is so small, we feel it safe to neglect extinction in our results.

\begin{figure*}
\includegraphics[width=\linewidth]{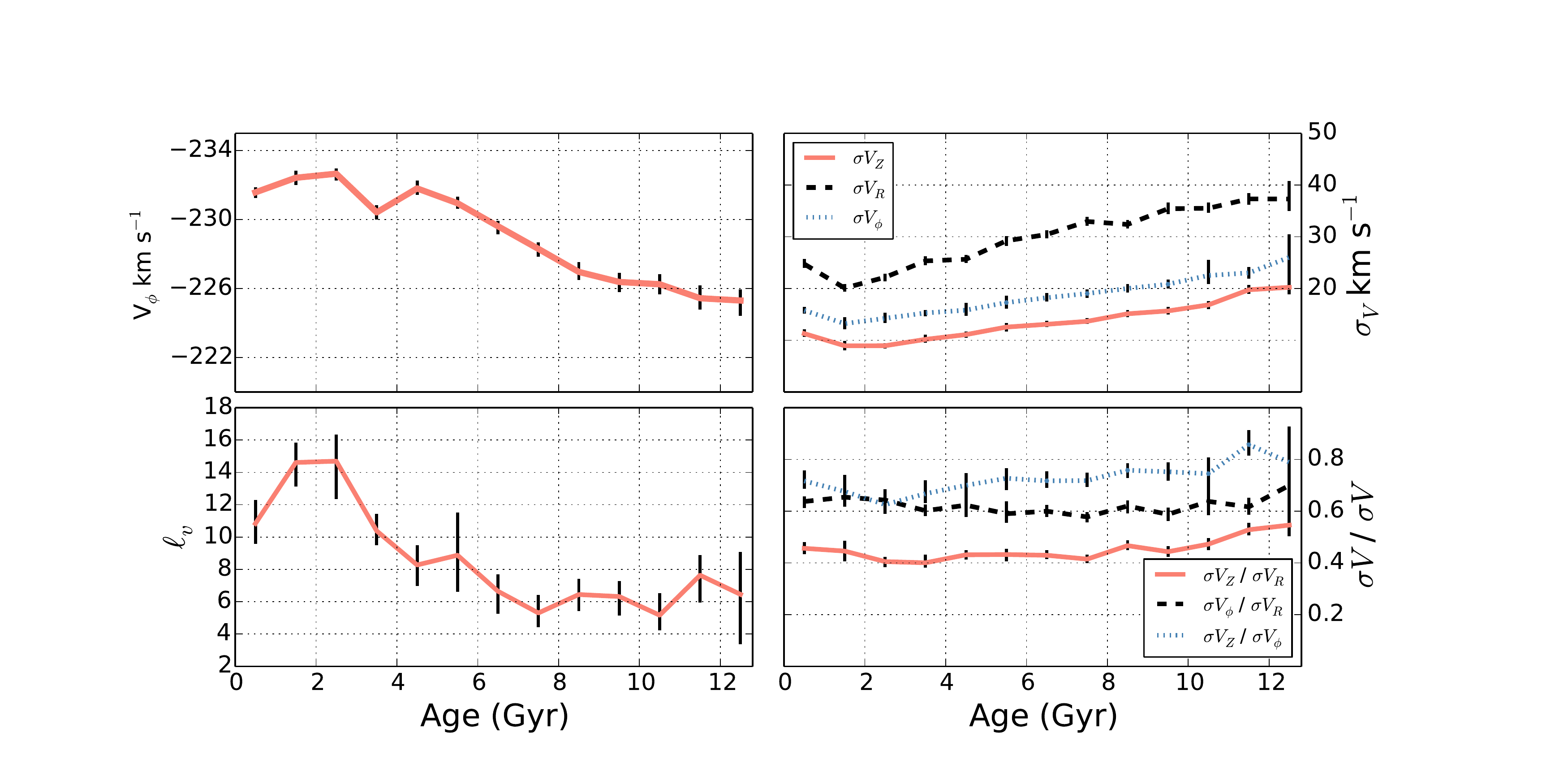}
\caption{\emph{Bottom left}: The vertex deviation of the velocities from the dynamical symmetry (equation 23 of \citealt{deh1998}). The younger stars show a larger deviation than the older stars, implying that their kinematics are largely determined by asymmetries in the disk. Over time, scattering increases the dispersion of these populations (\emph{top right}), and the influence of the velocity vertex decreases. In the \emph{bottom right} panel we show the ratios of the velocity dispersions. The \emph{top left} panel shows the average V$_{R}$ and V$_{\phi}$ velocities as a function of age. V$_{\phi}$ is seen to fall with age, which is likely an effect of the asymmetric drift (populations with higher velocity dispersions have lower V$_{\phi}$) and inside-out formation (older stars, being concentrated in the central regions, sacrifice more angular speed as they are heated out to the solar neighborhood). The errors are the 68\% confidence intervals, calculated by bootstrapping. The velocity dispersions and vertex angle estimations have had the average measurement errors subtracted, to reflect the underlying distributions. Three $\sigma$ velocity outliers have been iteratively removed from each bin.}
\label{fig:vertex_angle}
\end{figure*}

\section{Coordinates and Orbit Estimation}\label{sec:coords}

The coordinates used in this work are all right handed. The Sun is located at $(X, Y, Z) = (-8.27, 0, 0)$ kpc, and rotates at $(V_{R}, V_{\phi}, V_{Z}) = (0, -236, 0)$ km s$^{-1}$  (where $V_{R}$ is positive outward) -- or equivalently $(U, V, W) = (0, 236, 0)$ km s$^{-1}$. The solar motion with respect to the local standard of rest is: $(U, V, W) = (13.0, 12.24, 7.24)$ km s$^{-1}$. These values are all taken from \citep{sch2017, sch2012}.

The observables are transformed into XYZ space using the niave distance estimation of $d\ =\ 1/\pi$ and into UVW space using the method described in \citet{joh1987}\footnote{https://idlastro.gsfc.nasa.gov/ftp/pro/astro/gal\_uvw.pro}. The average values of, and errors on, these derived quantities are calculated by performing 100 random samples of the data scattered about the radial velocity, proper motion, and parallax errors. In this paper when we refer to velocities or positions, it is to the average values of these Monte-Carlo samples.

Orbits are calculated using the $galpy$ code described in \citet{bov2015}\footnote{https://github.com/jobovy/galpy}. The Milky Way potential used is the included MWPotential2014 (see section 3.5 of \citealt{bov2015}) which consists of 3 components: a power law bulge which is exponentially cut off \citep{mcm2011}; a Miyamoto-Nagai disk \citep{miy1975}; and an NFW halo \citep{nav1997}. The default potential is modified to be consistent with our adopted location, rotation speed, and peculiar solar velocity relative to the local standard of rest. Orbits are calculated on time steps such that the total energy change is negligible over the course of the orbit (the relative change in energy is less than $1 \times 10^{-9}$ for all but three of our orbits).

\section{Dynamics}\label{sec:dynamics}

\subsection{Vertex Angle and Velocity Dispersion}

In an axisymmetric potential, it is expected that stars will be born on near-circular orbits, and, as they are heated onto more eccentric, epicyclic orbits, the V$_{\phi}$ vs V$_{r}$ velocity plane will be symmetric about V$_{r}$. The Milky Way potential, however, is not axisymmetric, so the effects of the bar and the spiral arms can be quite influential in the structure of the velocity space and the velocity plane is consequently asymmetric.

For example, the well known Hercules stream may be stars trapped in orbits around Lagrange points of the bar \citep{per2017}; or, alternatively stars in the Outer Lindblad resonance \citep{hun2017}. The outer Lindblad resonance of the bar is also predicted to create observable deviations and bimodalities in the velocity plane \citep{deh2000}.

Besides the bar, the relative locations of the spiral arms are thought to affect the orientation of the ridgeline of the velocity plane \citep{ant2011}. The deviation of the velocity vertex (\citealt{deh1998}, equation 23),

\begin{equation}\label{eqn:vertex_deviation}
\ell_{v} \equiv \frac{1}{2} \text{arctan} \left( \frac{2 \sigma^{2}_{xy}}{\sigma^{2}_{xx} - \sigma^{2}_{yy}} \right),
\end{equation}

where $\sigma^{2}_{ij}$ are elements of the velocity tensor,

\begin{equation}\label{eqn:sigij}
\mean{(V_{i} - \mean{V_{i}}) \cdot (V_{j} - \mean{V_{j}})},
\end{equation}

describes this global tilt in the velocity plane.

It is expected that stars will be born on near-circular orbits with very little velocity dispersion relative to their siblings.
Stars are generally thought to be born on cool orbits with very little velocity dispersion relative to their siblings. Over time, they will be heated by interactions with molecular clouds (see, for example: \citealt{aum2016}, \citealt{gus2016}) and scattered by interactions with the non-axisymmetric potential;  and so their velocity dispersions will increase over time. Observationally, this means that younger stars will tend to be more concentrated in the velocity plane, while older stars will tend to be more diffuse.

In Figure \ref{fig:vertex_angle} we plot the velocity dispersions, ratios thereof, the vertex angle deviation, and rotational velocity curve. The velocity dispersions have the observational errors subtracted, and the vertex angle is corrected for the observational covariances as in \citet{smi2009}. We see the characteristic rise in all components of the velocity dispersion as a function of age, an expected result of heating over time (or, possibly, of the cooling of the source gas of the disk stars over time, as explored in \citealt{aum2016}). As the dispersions rise, we see the characteristic fall in the rotation speed, V$_{\phi}$, that is associated with the asymmetric drift. V$_{R}$ (not shown) does not change significantly with age, the average being -3.9  km s$^{-1}$, comparable to \citet{wil2013}.

The dispersion ratios indicate the relative importance of various heating mechanisms. The radial and azimuthal velocity dispersions are thought to be mostly increased by the spiral arms \citep{sel1984}, and so we expect them to rise proportionally -- as is seen by the flat black curve in the bottom right panel. The vertical velocity dispersions are thought to arise from interactions with molecular clouds in tandem with the spiral arms (\citealt{lac1984}, \citealt{jen1992}). It is interesting to see that the blue and red curves are largest for the oldest stars, as this implies one or more of the following: that vertical heating is more effective than planar heating; that older stars were born into vertically hotter orbits than today's young stars; or that mergers were more effective at heating the disk in earlier epochs.

\begin{figure*}[t!]
\includegraphics[width=\linewidth]{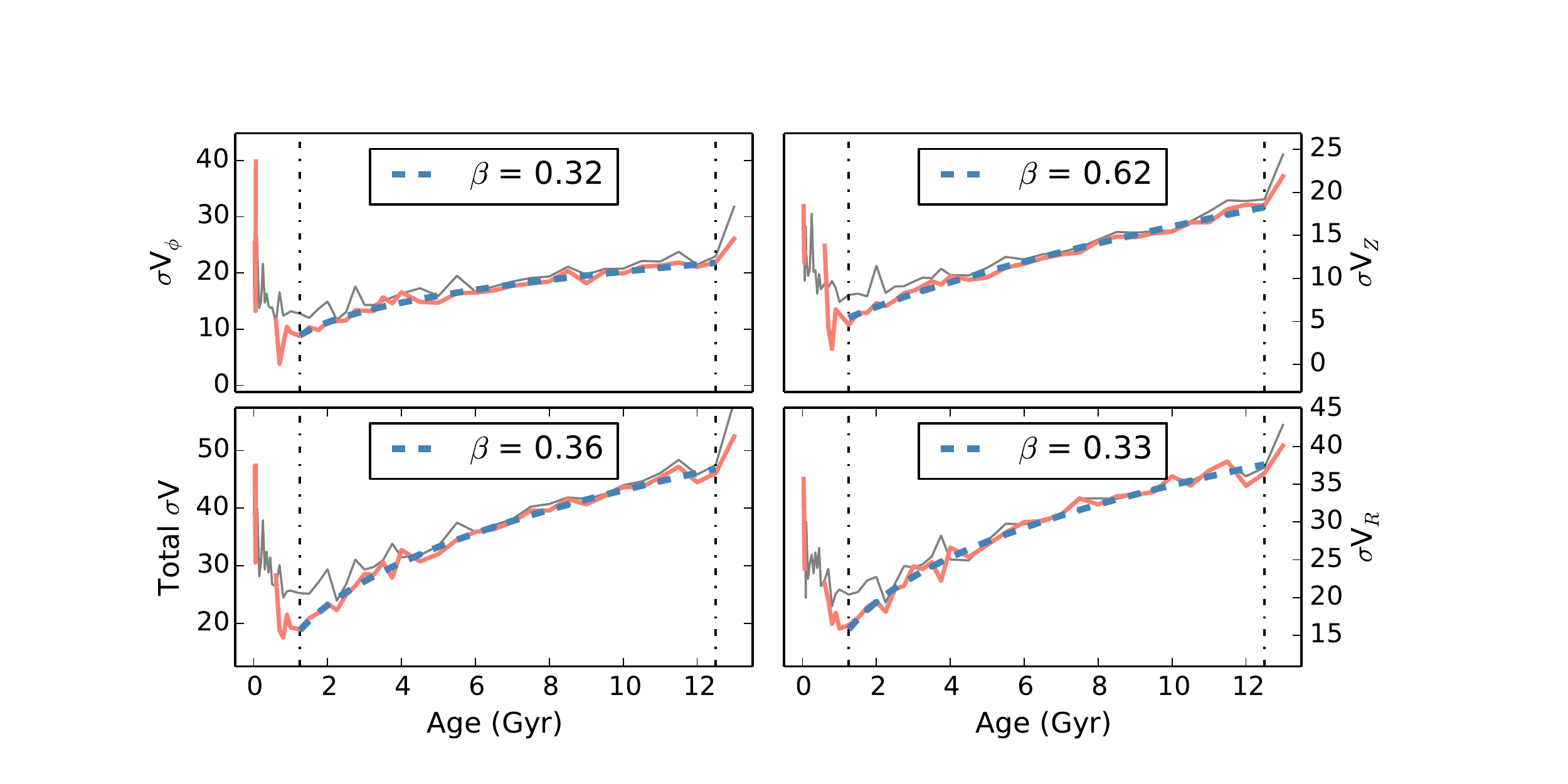}
\caption{The  velocity dispersions at each point in our age grid. All data are shown in the grey curves, while data with less than 100\% age errors, and with 3$\sigma$ velocity outliers removed, are shown in red. Bins younger than 1.25 Gyr (left dash-dotted line) all have less than 100 stars in them when making this selection (with the exception of the edge bins 0.03 and 0.04 Gyr). The oldest bin increases dramatically in all dispersion spaces, this is likely due to thick disk contamination, and possibly lower quality observations (as this is an edge-case bin). The number of counter-rotating stars is small, so we don't expect the halo to be a contributing factor. Fitting the cleaned data (with the average errors subtracted) results in the dashed-blue power-law fits. The data have been corrected for the observational errors and should reflect the true dispersions of the populations in each bin. Note that the fits are similar if we cut at 50\% age errors instead.
}
\label{fig:total_dispersion}
\end{figure*}

The vertex angle as a function of age has an interesting history. It was noted by \citet{par1958} that the vertex angle of stars changed sharply around a spectral class of F7. \citet{deh1998} note that this change should occur at the color where the main sequence lifetime of a star is the age of the Galactic disk, because redward of this break, stars of all ages are being observed. In our data, we are able to look at the vertex angle directly as a function of age, without the need for color as a proxy for age. We find that the younger stars -- less than $\sim$3 Gyr of age -- have a stronger deviation from symmetry in the velocity plane, and that this deviation drops from 3 Gyr until about 6.5 Gyr of age. Older than 6.5 Gyr of age, the vertex angle remains constant, but non-zero. This could imply that mixing and blurring play a more important role in the kinematics of stars after 6.5 Gyr of age, while when younger than 6.5 Gyr, the spiral arms, which heavily influence the vertex angle deviation, are more influential. We will look at this further in Section \ref{sec:age-met}.

\subsection{Ages and Dispersions}

In Figure \ref{fig:vertex_angle}, it can be seen that the very youngest age bin reverses the trend of positive $\partial \sigma_{V} / \partial \tau$ and negative $\partial \ell_{V} / \partial \tau$. Such a reversal has also been noticed in the bluest stars of \citet{aum2009} and the most metal rich stars \citet{ang2017}. An uptick in the $\sigma$V$_{\phi}$  distribution for the bluest stars is also seen in \citet{deh1998} who posit that the very youngest stars are probably not yet well mixed and travel in moving groups on orbits similar to their progenitor clouds.

When we attempt to fit our dispersions to a power law of the form

\begin{equation}\label{eqn:velocity_dispersion}
\sigma(x) = \sigma_{10} \left( \frac{x+x_{1}}{10 Gyr + x_{1}} \right)^{\beta},
\end{equation}

as in \citet{aum2009}, we obtain poor fits because of this uptick.

In Figure \ref{fig:total_dispersion} we plot the individual and total dispersions using our isochrone grid ages as bins, instead of the large bins used in Figure \ref{fig:vertex_angle}, so that we can more clearly investigate this uptick.

We see a jump in dispersion for stars in our old-age edge, which is possibly due to thick disk or halo contamination. The number of counter-rotating stars (which should most probably be halo stars, and should constitute roughly half of the halo contaminants in the sample) is negligible, about 0.15$\%$; so we expect that this uptick is most likely due to thick disk stars in our sample falling predominantly into this bin.

For stars younger than 1 Gyr, the velocity dispersions are higher than expected, the dispersions then flatten out until about 2 Gyr, at which point they start to increase with age. This goes against the canonical scenario of stars being predominantly born on cool, circular orbits and then being heated onto more eccentric ones. It could indicate that the younger stars are an unmixed population traveling in moving groups along the paths of their parent clouds (as proposed by \citealt{deh1998}).

When we consider only stars which have relative age errors less than 100\%, and also implement an iterative, three-sigma clipping procedure, we find that a large portion of the stars younger than $\sim$1.25 Gyr are removed. The only well populated bin below this cut is at 0.04 Gyr, which is adjacent to the removed edge bin of 0.03 Gyr. Since the majority of the stars contributing to the anomalous behavior in the young age bin are removed by more stringent quality cuts, or abut the edge of our age-grid, we treat them skeptically. With this in mind, we fit the velocity dispersions of stars with less than 100\% age errors with velocity outliers removed in the age range of 1.25 to 12.5 Gyr and find:

\begin{alignat}{3}
	\beta_{Tot},\ \sigma_{Tot, 10} &= (0.36,\ 43.1 \text{ km s}^{-1}), \nonumber\\
	\beta_{R},\ \sigma_{R, 10} &= (0.33,\ 34.9 \text{ km s}^{-1}), \nonumber\\
	\beta_{\phi},\ \sigma_{\phi, 10} &= (0.32,\ 20.2 \text{ km s}^{-1}), \nonumber\\
	\beta_{Z},\ \sigma_{Z, 10} &= (0.62,\ 16.0 \text{ km s}^{-1}), \label{eqn:betas}
\end{alignat}

These fits are performed accounting for the observational errors, and should reflect the underlying, intrinsic dispersions of the population. This $\beta_{Tot}$ value is similar to the values in \citet{bin2000} and \citet{aum2009} who found preferred values of 0.33 and 0.35, respectively. The individual components are well within the ranges found in the simulations of \citet{aum2016}. \citet{hol2009} -- who similarly exclude their youngest age bins (which have higher velocity dispersions than the slightly older bins) and oldest age bins -- find similar, but slightly higher, values for all components, with the exception of $\beta_{Z}$.

\subsection{The Oort Constants}

Other commonly used representations of disk dynamics are the so called Oort constants (\citealt{oll2003}, \citealt{oor1927}):

\begin{alignat}{3}
	2A & =  &&\frac{\mean{V_{C}}}{R_{0}} &&- \mean{V_{C,R}} - \frac{\mean{V_{R,\phi}}}{R_{0}}, \nonumber\\
	2B & = -&&\frac{\mean{V_{C}}}{R_{0}} &&- \mean{V_{C,R}} + \frac{\mean{V_{R,\phi}}}{R_{0}}, \nonumber \\
	2C & = -&&\frac{\mean{V_{R}}}{R_{0}} &&+ \mean{V_{R,R}} - \frac{\mean{V_{C,\phi}}}{R_{0}}, \nonumber \\
	2K & =  &&\frac{\mean{V_{R}}}{R_{0}} &&+ \mean{V_{R,R}} + \frac{\mean{V_{C,\phi}}}{R_{0}}. \label{eqn:oorts}
\end{alignat}

\begin{figure}
\includegraphics[width=\linewidth]{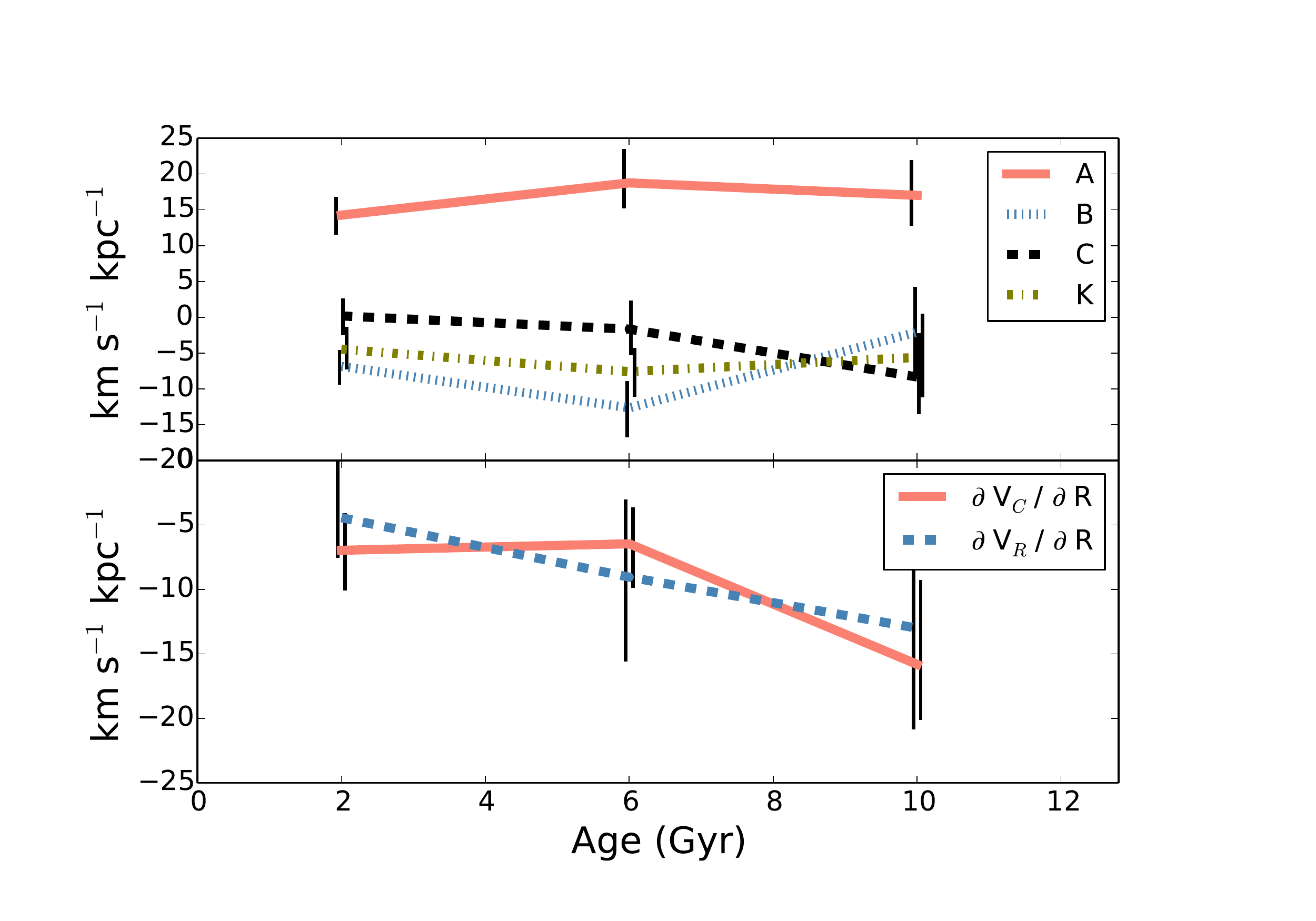}
\caption{The Oort constants and 68\% confidence intervals calculated for our data, via bootstrapping, as a function of their derived ages. Also plotted are the slopes of the rotation curves, $\partial V_{C} / \partial R$ and $\partial V_{R} / \partial R$. Note that the errorbars are shifted by 0.05 Gyr to avoid overlap.}
\label{fig:oort}
\end{figure}
\begin{table}
\begin{center}
\caption{Oort Parameters as a Function of Age}
\begin{tabular}{|c|c|c|c|}
\tableline
 & Young & Intermediate & Old \\
\tableline
\tableline
\tableline
A & 14.3 & 18.8 & 17.0 \\
\tableline
B & -6.7 & -12.7 & -2.1 \\
\tableline
C & 0.2 & -1.7 & -8.3 \\
\tableline
K & -4.5 &-7.6 & -5.6 \\
\tableline
$\frac{\partial V_{R}}{\partial R}$ & -4.4 & -9.0 & -13.0 \\
\tableline
$\frac{\partial V_{C}}{ \partial R}$ & -7.0 & -6.5 & -15.8 \\
\tableline
\end{tabular}
\end{center}
\label{tab:oort}
\end{table}

\begin{figure}
\includegraphics[width=\linewidth]{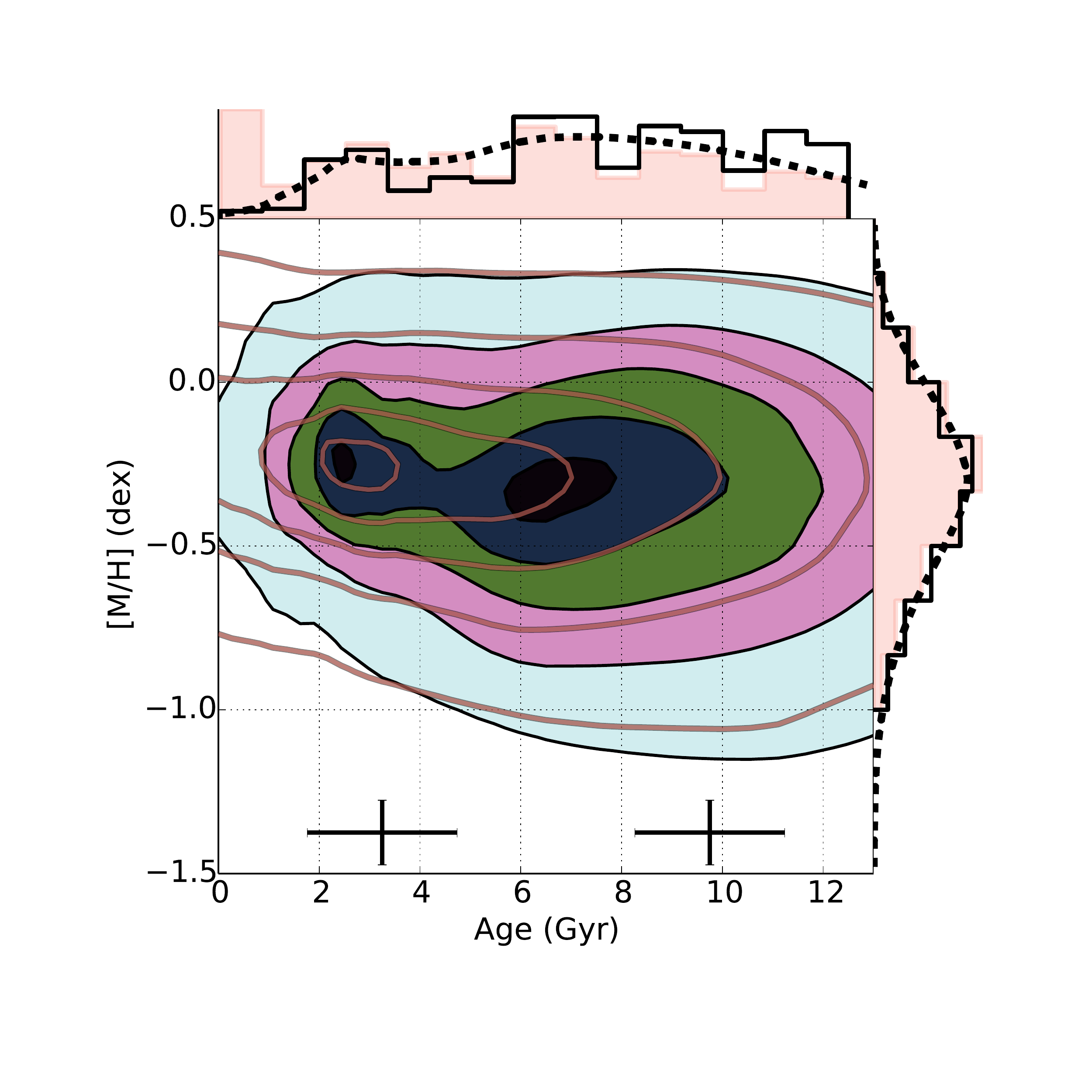}
\caption{The age-metallicity relation of our sample. The distribution of age estimates given by the method outlined in Section \ref{sec:age_estimation} after having been subjected to the cuts in the same Section and weighted by the procedure outlined in Appendix \ref{app:sel_fn}. The data are spread over the PDFs of their age estimates and the PDF of the metallicity estimate and error. The contours indicate the 5\%, 25\%, 50\%, 75\% and 95\% levels of the maximum density. \\
The red contour lines and shaded histograms indicate the distribution of the entire sample (peaking around 3 gyr of age). The filled-color contours and black unshaded histograms are the distributions of the sample with less than 50\% errors on the age estimates. The black-dashed distributions on the axes are the probability density distribution (i.e. smoothed by the uncertainties) of this higher precision subset. Average error bars for low and high age stars for the higher precision subset (filled contours) are shown.}
\label{fig:age_histogram}
\end{figure}

The notation $\mean{V_{R}}$ is used for the mean value of the radial velocity of the population. Since our right-handed coordinate system means that $V_{\phi}$ is negative for disc rotation, here we adopt the notation $V_{C}$ = -$V_{\phi}$ for azimuthal velocity. The additional subscripts ,$\phi$ and ,$R$ indicate derivatives with respect to Galactocentric angle $\phi$ and radius $R$. These constants contain information about the motion of the Galaxy, such as the angular rotation speed and the slope of the velocity curve. Since we have spectroscopy (radial velocities) for our data, we may estimate these values directly, but for comparison with the literature, we will also calculate these Oort constants in different age bins.

Figure \ref{fig:oort} shows our findings for the calculated Oort constants, as well as associated velocity derivatives $\partial V_{C} / \partial R$ and $\partial V_{R} / \partial R$, as a function of time. Our average values for young, intermediate age, and old populations are shown in Table 1.

\citet{bov2017} found (A, B, C, K) = (15.3, -11.9, -3.2, -3.3) km s$^{-1}$ kpc$^{-1}$ using main sequence stars from TGAS; \citet{oll2003} estimated (A, B, C) = (16, -17, -10) km s$^{-1}$ kpc$^{-1}$  using Tycho-2 and ACT data \citep{urb1998}; \citet{fea1997} calculated (A, B) = (14.82, -12.37) km s$^{-1}$ kpc$^{-1}$ using Cepheids.

Our value for $\partial V_{R} / \partial R$ (which is equivalent to C+K) grows steeper as a function of age and our value for $\partial V_{C} / \partial R$ (which is the same as -A-B) is steepest for the oldest age bin. We find these slopes to be negative for all ages, and most negative for the oldest stars. Our values for $\partial V_{R} / \partial R$ are in agreement with the literature (-6.5 km s$^{-1}$ kpc$^{-1}$  according to \citealt{bov2017} or -4 to -10 km s$^{-1}$ kpc$^{-1}$  from \citealt{sie2011}). Our $\partial V_{C} / \partial R$ values are larger than what are generally found (-3.4 \citealt{bov2017}, -1 to -13 \citealt{sie2011}, -2.4 \citealt{fea1997}, 1 \citealt{oll2003}, -4 \citealt{hua2016}; in units of km s$^{-1}$ kpc$^{-1}$).

\section{Abundances}\label{sec:abundances}

\subsection{Age-Metallicity Relation}\label{sec:age-met}

\begin{figure*}
\includegraphics[width=\textwidth]{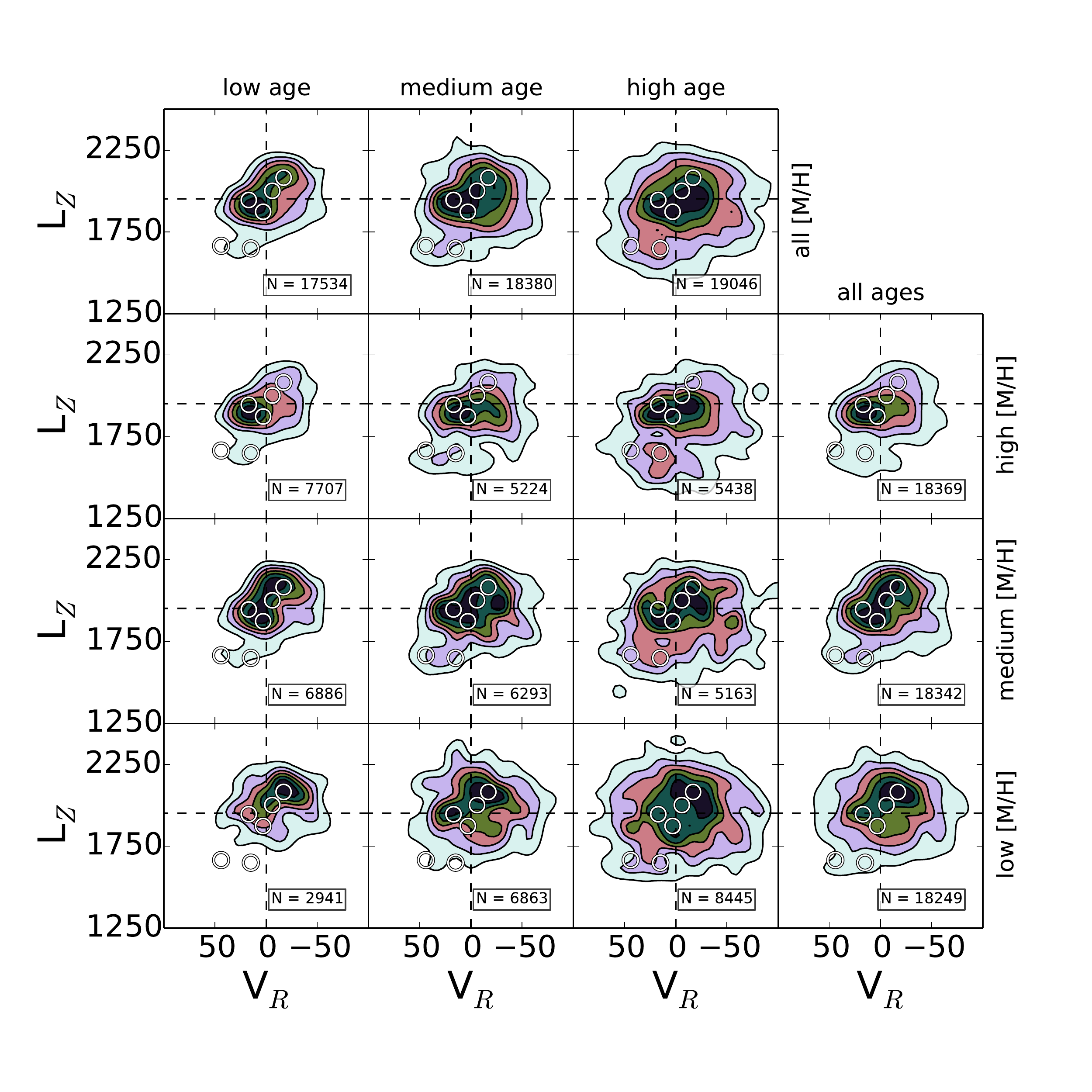}
\caption{The angular-momentum versus velocity plane as a function of age and metallicity; in each frame there are six contours, smoothed through a cubic spline interpolation, equally spaced between zero and the maximum density. We can see that as age increases, a strong concentration along the vertex gives way to a more symmetric distribution. This is commonly attributed to heating over time, but can also reflect a different birth environment (when the vertex deviation angle was different, as it is suggested that this angle is a function of position relative to non-axisymmetries in the disk; \citealt{ant2011}).\\
Also visible is the clear preference for higher metallicity stars to be rotating slower than the local standard of rest, and for lower metallicity stars to be rotating faster. This is a consequence of blurring and is visible, to varying extents, for all ages. \\
The bins are chosen to approximately trisect the data in age ( $<$ 6 Gyr, 6 to 9 Gyr, $>$ 9 Gyr), and, separately in [M/H] ( $<$ -0.45 dex, -0.45 to -0.21 dex, $>$ -0.21 dex).
\\
The overplotted circles indicate the locations of the moving groups (from highest to lowest $L_{Z}$; locations taken from Table 2 of \citealt{ant2012}): Sirius, Coma Berenices, Hyades, Pleiades, and two detections of Hercules. The Hercules moving group is especially apparent in the high-metallicity, high-age panel.
}
\label{fig:velocity_grid}
\end{figure*}

In Figure \ref{fig:age_histogram} we plot the age-metallicity distribution of our sample of stars. The age distribution of our sample peaks at around 3 Gyr, when factoring in the age probability spreads. \citet{feu2016} found an age distribution peaked at around 3-4 Gyr using giants in APOGEE and Hipparcos; \citet{cas2016} found astroseismic ages for giants observed by Kepler \citep{bor2010} to have a double peak at around 2 Gyr and 4 Gyr; \citet{nor2004} find ages peaked around 2 Gyr for F and G dwarfs in the smaller volume of the Geneva Copenhagen Survey sample.

To test whether our ages are biased by degeneracy between metallicity and the estimated ages, we can utilize a secondary effect of the ``blurring'' mechanism which increases the velocity dispersion of similarly aged stars over time. Since the gradient $\partial [M/H] / \partial R$ of the interstellar medium is negative, stars born at smaller radii are typically more enriched than stars born at larger radii \emph{at the same time}. So, when looking a collection of stars in a small volume, around the same radius, at the same age, the stars rotating more rapidly tend to be more metal poor, and the stars rotating more slowly tend to be more metal rich. 

Figure \ref{fig:velocity_grid} examines the angular momentum ($L_{Z}$) velocity plane as a function of both age and enrichment, it can be seen that these two attributes have very different effects on our sample. In terms of age, we can easily see the effects of heating over time: epicyclic heating causes the velocity dispersion to increase, and simultaneously the tilt of the velocity vertex becomes less and less apparent.

In terms of metallicity, we can see the effect of the $\partial [M/H] / \partial R$ of the disk ISM on the velocities of similarly aged stars in our sample. The lower metallicity stars (on larger guiding center radii orbits) tend to be rotating faster than the higher metallicity stars observed in our volume, caused by blurring and the conservation of angular momentum. Hopefully Figure \ref{fig:velocity_grid} puts to rest some of the worry about possible age-[M/H] degeneracies, as the two spaces are seen to have distinct and differing effects on our sample.

\citet{ant2017} noted that, for their thin disk objects, the more quickly orbiting stars had generally lower metallicities than the stars which were moving with slower V$_{\phi}$ speeds, as is expected from blurring. They also see that the stars in their data which were traveling inward (with negative V$_{R}$) in the galaxy had a systemically higher rotational velocity than stars which were traveling outward in the galaxy. This is a natural consequence of the vertex angle deviation. This is all shown nicely in Figure \ref{fig:velocity_grid} with metal rich objects moving outward and at lower L$_{Z}$, and vice-versa for metal poor objects.

We also see the presence of numerous moving groups in the data. Hercules, for example, while standing out most prominently in the older, higher metallicity populations, is seen faintly in most of the panels. Being present at a wide range of ages and metallicities implies that Hercules is a resonant feature. It is noticeably weaker for younger stars with low metallicity. However, comparing age distributions is difficult due to survey selection effects, such as the volume under consideration or the type of stars being measured (see, for example, section 3 of \citealt{aum2016}).

\subsection{Inside Out Enrichment}

\begin{figure}
\includegraphics[width=\linewidth]{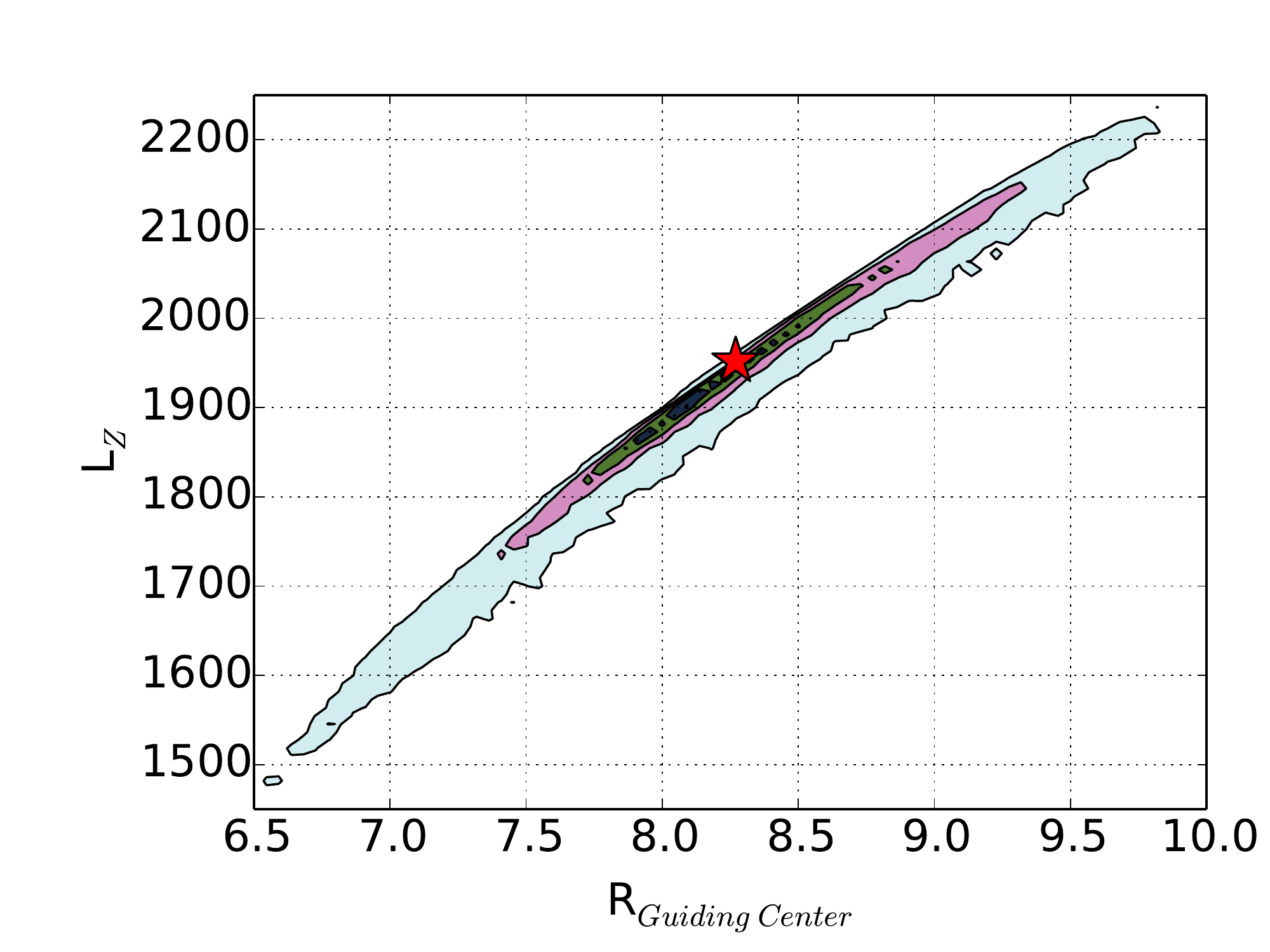}
\caption{The relationship between a stars observed angular momentum, $L_{Z}$, and its inferred guiding center (birth) radius based on orbit integrations. The contour levels are the 5\%, 25\%, 50\%, 75\% and 95\% of the maximum. The two are well mapped to each other and are relatively monotonic. Because of this, we use $L_{Z}$ as a proxy for birth radius for the rest of the analysis. The red star is at the assumed angular momentum of the local standard of rest, and current radius.}
\label{fig:vphi_rg}
\end{figure}

If the rotation curve is relatively flat, we may think of L$_{Z}$ as a proxy for the Galactocentric guiding center radius of the orbit, which should be the birth radius in the absence of churning. In Figure \ref{fig:vphi_rg} we show the relationship between the angular momentum of stars in our sample and the guiding center radius estimated from the stars' orbits. While not linear, the two are fairly well and monotonically correlated. For the rest of the paper we will use  $L_{Z}$ as a proxy for this guiding center radius, removing any assumptions implied by the potential while retaining the basic information about a star's birth radius.

\begin{figure}
\includegraphics[width=\linewidth]{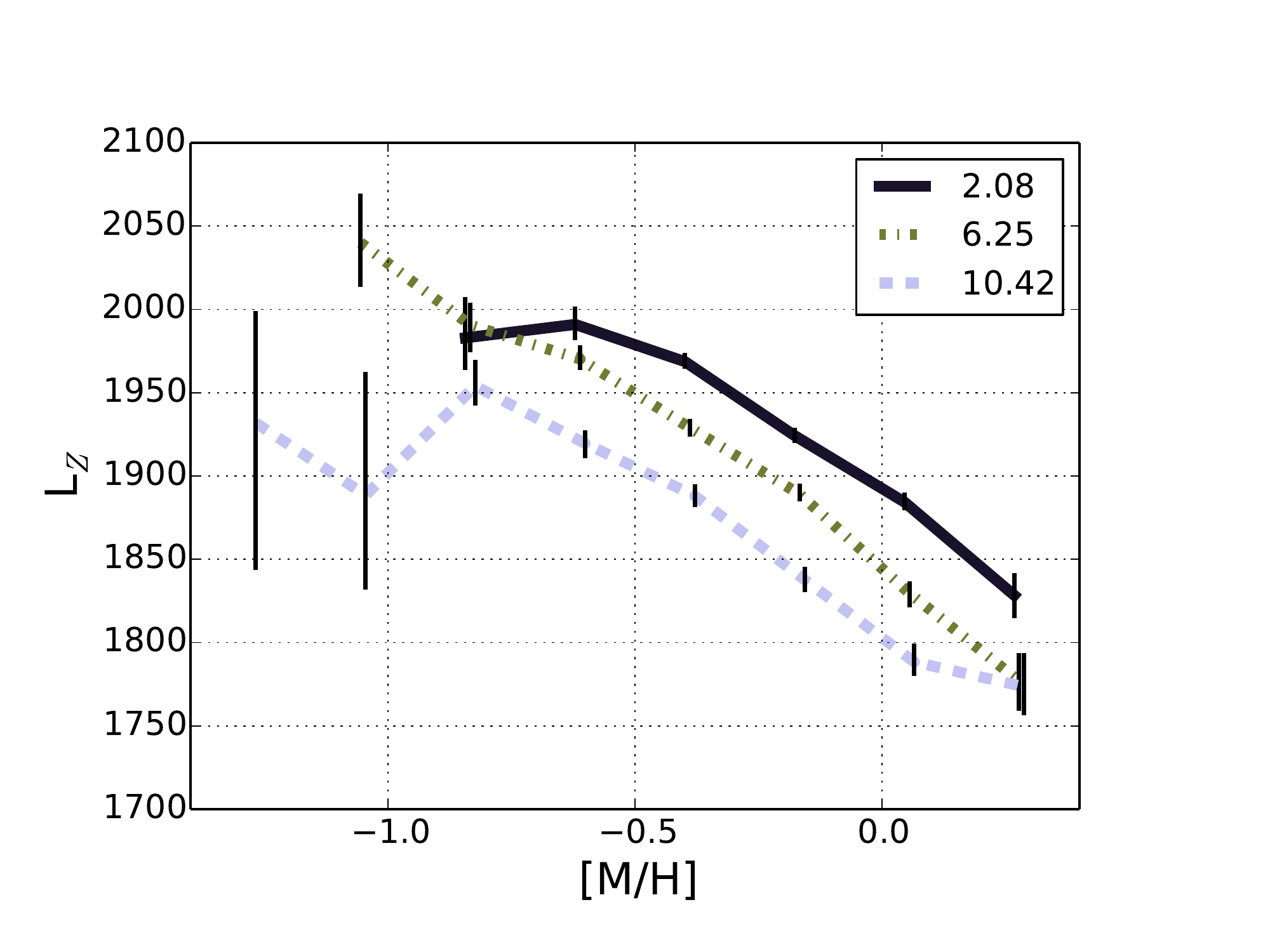}
\caption{Angular momentum is plotted as a function of metallicity in independent bins spaced $\sim$0.2 dex apart for equidistanced age populations. Only bins with 10 or more successful Monte-Carlo samplings, which each have 50 or more data points, are shown, and the error bars indicate the 68\% confidence intervals of those 10 or more samplings. The negative slope of the line arises from the vertex angle of the local velocity field, which is caused by more metal poor stars, originating farther outward in the Galaxy, visiting the solar neighborhood at the perigalacticon of their epicyclic orbits, hence they are traveling with greater angular momentum (and vice-versa for the metal rich stars). The movement of this line to the right with decreasing age is a consequence of enrichment happening at all radii (if we use $L_{Z}$ as a proxy for guiding center radius of the stars). Note that the error-bars are offset to avoid overlapping.}
\label{fig:vphi_met_age}
\end{figure}

\begin{figure}
\includegraphics[width=\linewidth]{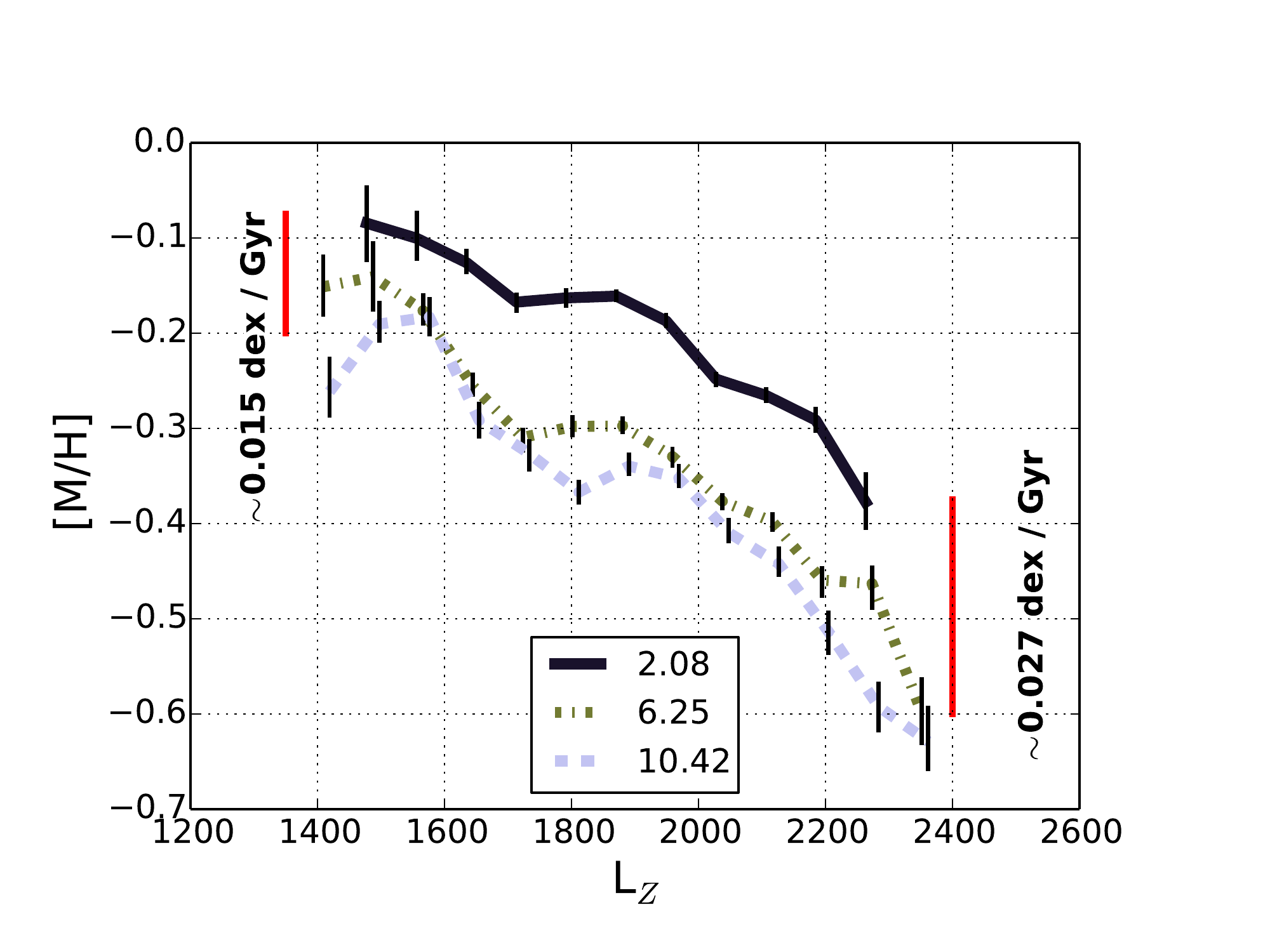}
\caption{Similar to Figure \ref{fig:vphi_met_age} but with the axes reversed. Using $L_{Z}$ as a proxy for birth radius, we see that the interior regions (lower angular momentum) are more enriched for any given age (the negative slope in each individual line). For the interior regions, the different aged populations are more similar to each other than those respectively aged populations in the outer regions.
}
\label{fig:met_vphi_age}
\end{figure}

Examining trends in enrichment of populations as a function of age can lead to insights into the formation processes of the Galaxy. In Figure \ref{fig:vphi_met_age} we plot the angular momentum as a function of metallicity and age. We notice that older stars are rotating with less angular momentum than younger stars at the same metallicity. We also see that, for a given age, higher metallicity stars are orbiting with less angular momentum than lower metallicity stars (as seen in, for example, \citealt{lee2011}). Since the angular momentum is broadly representative of the birth radius of a star (in the absence of churning), this means that: 1) older stars of a given metallicity are born more interior in the Galaxy than younger stars of the same metallicity, and 2) more metal rich stars of a given age are born more interior to more depleted stars of the same age. These are both expectations of an inside-out formation scenario.

In the oldest age bin, at the metal poor end, there is a downtick to lower angular momenta. This could possibly be explained by contaminating thick disk stars.

We may also switch the axes to look at metallicity as a function of angular momentum and age (Figure \ref{fig:met_vphi_age}). In this figure, using $L_{Z}$ as a proxy for radius, we see that at all radii, metallicity increases with age, i.e. $\partial [M/H] / \partial \tau$ is positive. We also see that for any given age $\partial [M/H] / \partial R$ is negative. However, these slopes are different for different ages, with there being more of a metallicity difference between the inner portions of the disk and the outer portions of the disk for old stars; and less of a difference for young stars ($|\partial [M/H] / \partial R|_{\tau=10} > |\partial [M/H] / \partial R|_{\tau=1}$). Another way to say this is that $(\partial [M/H] / \partial \tau)_{Outer Disk} > (\partial [M/H] / \partial \tau)_{Inner Disk}$, in Figure \ref{fig:met_vphi_age}, for example, the inner disk enriches at a rate of about 0.015 dex Gyr$^{-1}$, while the outer disk enriches at a rate of about 0.027 dex Gyr$^{-1}$ (between the 2.08 and 10.42 Gyr old populations). This again supports inside out formation, with the interior regions reaching high levels of enrichment on a smaller timescale than the outer regions.

\subsection{Migration}

\begin{figure}
\includegraphics[width=\linewidth]{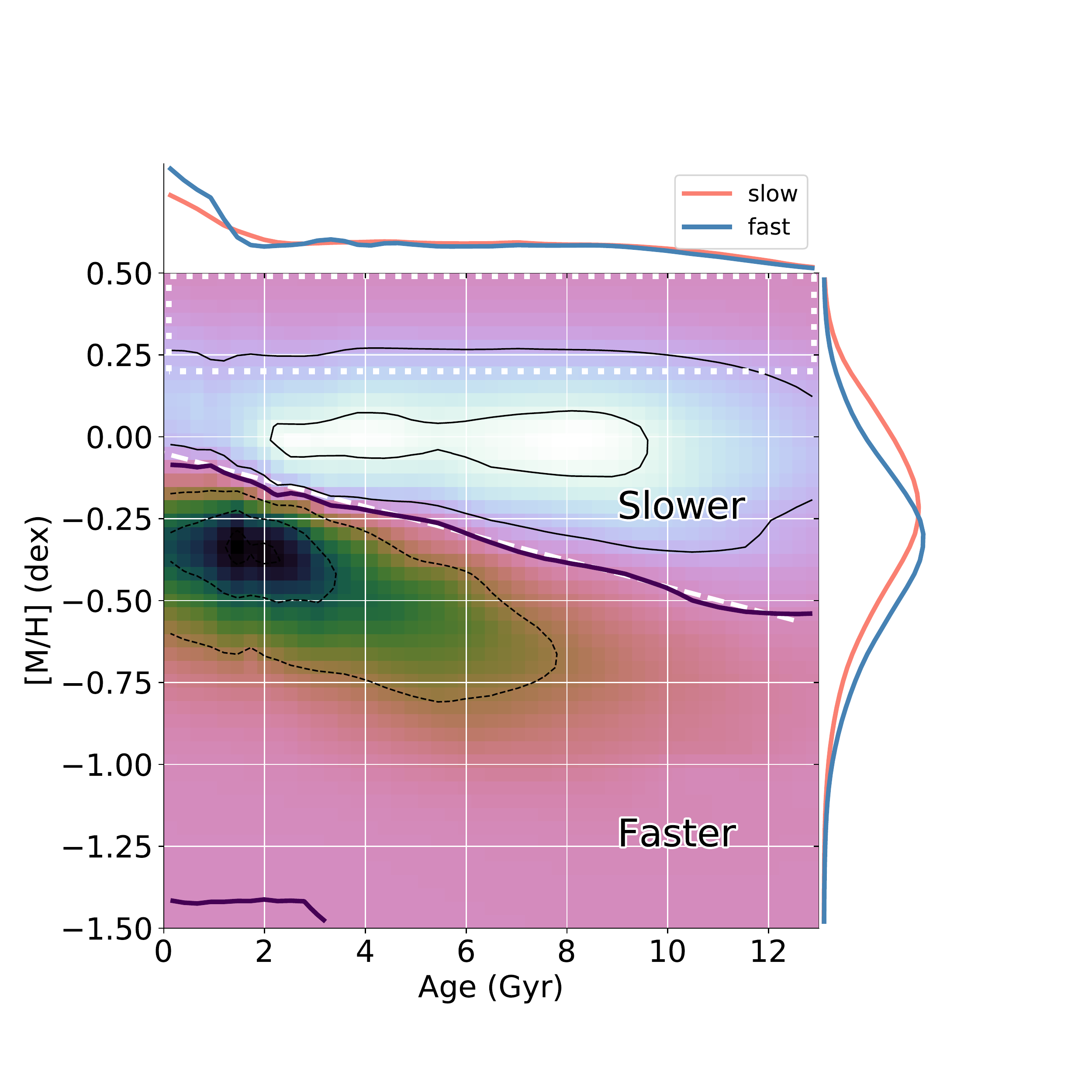}
\caption{
The age-metallicity relationship (as in Figure \ref{fig:age_histogram}) residual between slowly orbiting stars ($L_{Z}<$1952 kpc km s$^{-1}$, from the interior Galaxy, light areas) minus quickly orbiting stars (from the outer Galaxy, darker areas). The thick black contour indicates the zero-line (the other contours are at the 5\%, 25\%, 75\% and 95\% levels), and the underplotted white line could be thought of as the enrichment rate at the solar neighborhood (with a by-eye slope of $\partial [M/H] / \partial \tau$$\sim$0.04 dex Gyr$^{-1}$). The side panel shows the metallicity distributions. \\
The white dotted box indicates an area where the enrichment level is greater than the local ISM, and the upper histogram shows  the age distribution of stars in this white box which are orbiting slower than the Sun (having blurred out from interior regions which are more metal rich) and which are orbiting more quickly than the Sun (which can not have been heated to their current locations, since the exterior regions are not this enriched). The blue histogram therefore shows probably churned stars, and has a surplus at less than 2 Gyr of age compared to the other distribution. \\
All these data are smoothed over the full age PDF, the metallicity errors, and are then weighted by the selection function.
}
\label{fig:inout_residual}
\end{figure}

In a figure similar to Figure \ref{fig:vphi_met_age}, \citet{ant2017} found a peculiar behavior where the most metal rich stars ($>$+0.5 dex) in their sample were rotating more swiftly than their slightly more metal poor neighbors (from +0.1 dex to +0.5 dex). A possible interpretation of this is that these stars have had their angular momenta increased by way of churning. Since, for such metal rich objects to have travelled to our observational space, they will have needed to sacrifice a lot of angular velocity to retain the same angular momentum, as is the expectation if epicyclic heating (blurring) is the only mechanism at play.

While we see such behavior in our uncut sample, the combination of quality cuts we perform excises these stars from Figure \ref{fig:vphi_met_age} (the most efficacious being the removal of all data which does not fall within 3 $\sigma$ of its errors from an isochrone). As pointed out by Ralph Sch\"{o}nrich (private communication), these stars could simply be solar-metallicity stars that have been scattered into the high metallicity tail of the distribution due to large observational errors. In that case our 3$\sigma$ cut would remove these as their metallicities would not be consistent with our adopted isochrones.

In a separate study of migration from the inner disk, \citet[see also \citealt{kor2015}]{hay2017} calculated orbital parameters for Gaia-ESO \citep{gil2012} stars and found that a portion of those with iron abundances greater than +0.1 dex had orbital speeds inconsistent with being born in the solar neighborhood or farther interior, as implied by that level of enrichment\footnote{Note that the local interstellar medium is estimated to be less enhanced than about +0.2 dex based on the abundances of nearby, short-lived stars and clusters. The abundances of local, short lived O and B stars has been estimated to be between -0.07 and +0.03 dex (\citealt{nie2012}, \citealt{prz2008}), the abundance levels of red giants in open clusters around the solar radius is found to be between +0.1 and +0.2 dex \citep{fri2013}, and the iron abundance of APOGEE red giants in clusters around the solar radius is estimated to be less than +0.1 dex on average \citep{cun2016}. The estimates from the O and B stars is probably closer to the real enhancement rate of the local ISM, since they have shorter lifespans than the red giant samples and so are probably less affected by possible migration or recent enrichment.}. With a similar test, we find that 266 of 1053 (about 25\%) of stars in our sample with metallicities over +0.2 dex are rotating faster than the local rotation speed, indicating that they have guiding center radii outside the solar radius. The contradiction between kinematics implying that these stars have guiding center radii external to the solar neighborhood and abundances implying interior birth environments could be rectified by a churning scenario pulling these stars outward.

Interestingly, when we collect a sample of these metal rich, quickly orbiting stars (top panel of Figure \ref{fig:inout_residual}), we find their age distribution to be more weighted to younger ages than slowly rotating metal rich stars. If these are bona fide churned stars, the younger-skewed age distribution could be indicative of churning being a more effective migratory mechanism than blurring for young populations, or on short time scales.

Also in Figure \ref{fig:inout_residual}, we plot the age-metallicity relationship for the residual between stars with less angular momentum than the local standard of rest ($L_{Z}\approx1952$  kpc km s$^{-1}$) and stars with more angular momentum (the residual between two, 2-dimensional histograms, which each integrate to 1). In general, the boundary between the two populations is characterized by a line which gets gradually more metal rich as time passes, which would represent the enrichment of the solar neighborhood in the absence of churning (with a slope of about $\partial [M/H] / \partial R$ $\sim$ 0.04 dex Gyr$^{-1}$). The stars which are rotating with less angular momentum than the local standard of rest tend to be more metal rich than this local ``enrichment rate" line since they tend to have been born at smaller radii than the solar circle; and stars which are orbiting with more angular momentum tend to be more metal poor than this line. This is expected from the conservation of angular momentum as stars are heated onto eccentric orbits.

\section{Discussion and Conclusions}\label{sec:conclusions}

In this paper we have analyzed data from the TGAS-RAVE-LAMOST crossmatched and co-added data set. With directly measured parallaxes, instead of photometrically or spectroscopically inferred parallaxes, we are able to estimate the ages of these stars with greater precision than previous studies using bulk RAVE and LAMOST data alone.

Using a sample of $\sim$55,000 high confidence measurements, which pass a variety of quality cuts, we have calculated the velocity dispersion as a function of age in all three components. Our values are similar to those found by \citet{bin2000} using Hipparcos data, \citet{aum2009} using Hipparcos data combined with Geneva-Copenhagen Survey radial velocities, and \citet{hol2009} again with Geneva-Copenhagen Survey and Hipparcos data (see Equation \ref{eqn:betas}).

However, to obtain good fits it was necessary to omit the data in our very youngest and very oldest age bins. The oldest age bin, we expect, suffers from an amount of thick disk contamination, so the fact that it has more dispersion than expected is reasonable. Since our observational volume is relatively large, it is perhaps not surprising our sample includes thick disk stars. Other samples covering similar volumes have also seen an increasing dispersion (e.g. the Hipparcos giant sample of \citealt{feu2016}), whereas some more local samples have not (e.g. Geneva-Copenhagen survey study by \citealt{hol2009}). Another possible explanation for these different findings is that the different studies have different age error magnitudes. \citet{mar2014} showed that age errors of 30\% can effectively smooth and erase jumps in the age-velocity relation, as they see in their simulations with regard to an old and kinematically hot population.

The younger stars, which have an unexpectedly high velocity dispersion, are puzzling though. We note that \citet{hol2009} also omit their youngest aged ($<\sim$1.5 Gyr) stars from their velocity dispersion fit, and we can see that these stars have similar or even higher velocity dispersions than their slightly older neighbors. \citet{deh1998} also note that their very bluest stars (indicative of the youngest age group) do not obey the expected velocity dispersion relation. This could be caused by the star forming regions being on orbits which are not kinematically well-mixed (as noted in \citealt{deh1998}). There could also be contamination in these bins from, for instance: blue horizontal branch stars or blue stragglers, which may appear young; or multiple stars which would have additional components to their velocity dispersions and are more common in high-mass, shorter lived populations.

When we implement more severe quality cuts on the size of the relative age errors and iteratively remove velocity outliers, we find that the biggest contribution to this rising velocity dispersion is coming from the two youngest age bins, which could indicate that this higher dispersion is just an edge effect.

Another measure of the kinematics, the velocity vertex deviation angle, which is formed by local potential non-axisymmetries and is gradually erased for older populations by changing potential (e.g. as the solar neighborhood changes position with respect to the spiral arms over time) and gradual heating, shows a counterintuitive reversal for the very youngest stars as well. Again, this youngest age bin is possibly contaminated by incorrectly aged stars in the edge bins, and not reliable. Curiously, this reversal is also seen by \citet{aum2009} for their very bluest stars, and by \citet{ang2017} for their most metal rich and least alpha enhanced stars. In our data, the velocity vertex deviation angle stops changing noticeably for ages above about 6.5 Gyr of age. This could imply a timescale for a local population to become well mixed, no longer showing appreciable signs of the potential they were born in.

We have examined the relationship between rotational velocity (as a proxy for orbital radius) and metallicity. There is clear evidence for all radii to be enriching over time, with interior radii always being more enriched than outer radii, lending further support to inside-out formation hypotheses. However, we also note that the amount of enrichment over time is greater in the outer regions of the disk than the inner portions, about $\sim$0.225 dex  over the range of our data, compared to $\sim$0.125 dex. This would imply that the central regions formed and enriched quickly, while the outer regions have had a slower enrichment history starting at a later time.

We note that, for our sample of stars, the dividing line between stars rotating more slowly than the local standard of rest ($L_{Z}\approx1952$  kpc km s$^{-1}$) and stars rotating more quickly in the age-metallicity diagram -- which could be indicative of the enrichment rate of the solar neighborhood if the main method of migration in the Galaxy is blurring -- is at  $\partial [M/H] / \partial \tau$$\sim$0.04 dex Gyr$^{-1}$. This is fiducially similar to observations by \citet{feu2016} and \citet{ber2018}, as well as to model parameters in \citet{sch2009}. 

We have also examined the slope of the rotation curve and the radial velocity curve. We find them both to be falling as a function of radius, with the oldest stars exhibiting the steepest slopes of descent.

In the data of \citet{ant2017}, the highest metallicity objects could be seen to be rotating more swiftly than would be expected if they had been transported to the local volume by blurring. We do not see such behavior in our data in Figure \ref{fig:vphi_met_age}. However, following the logic of \citet{hay2017}, we count, in our data, 266 stars with metallicities greater than +0.2 dex which are rotating faster than the local rotation speed. This is out of a total of 1053 stars at this enrichment level or higher -- a rate of $\sim$25\% which is in good agreement with the lower bound on such stars of 20\% found in that study.

It is intriguing that the stars which could not have been born in the solar neighborhood (having metallicities greater than +0.2 dex), yet have solar neighborhood dynamics, have a younger age distribution than other stars at the same metallicity, with a surplus at less than 2 Gyr of age). We expect that these stars were moved to orbits with new angular momenta via churning, which is thought to be most effective on cooler, and hence younger, populations. While churning should affect all populations at various points in time, an excess is apparent in the young population. We suppose that blurring, which takes time to build up the radial extent of orbits and tends to flatten metallicity gradients, could spread out this excess on longer timescales.

\section{Acknowledgements}

We thank the developers and maintainers of the following software libraries which were used in this work: Topcat \citep{tay2005},  NumPy \citep{van2011}, SciPy \citep{jon2001}, AstroPy \citep{ast2013}, astroML \citep{van2014}, galpy \citep{bov2015}, padova\footnote{https://github.com/jonathansick/padova}, pytess\footnote{https://pythonhosted.org/Pytess/}, matplotlib  \citep{hun2007}, IPython \citep{per2007} and Python.

We thank Emma Small for invaluable insight, expertise and guidance with regards to isochrones. We also thank Jerry Sellwood, Wyn Evans, Ralph Sch\"{o}nrich, and Victor Debattista for helpful discussions on dynamics.

JJV gratefully acknowledges the support of a LAMOST fellowship, support from the Chinese Academy of Sciences President's International Fellowship Initiative, as well as NSFC grant 11650110439. M.C.S. acknowledges financial support from the CAS One Hundred Talent Fund and from NSFC grants 11673083 and 11333003. This work was also supported by the National Key Basic Research Program of China 2014CB845700.

Guoshoujing Telescope (the Large Sky Area Multi-Object Fiber Spectroscopic Telescope LAMOST) is a National Major Scientific Project built by the Chinese Academy of Sciences. Funding for the project has been provided by the National Development and Reform Commission. LAMOST is operated and managed by the National Astronomical Observatories, Chinese Academy of Sciences.

This work has made use of data from the European Space Agency (ESA) mission {\it Gaia} (\url{https://www.cosmos.esa.int/gaia}), processed by the {\it Gaia} Data Processing and Analysis Consortium (DPAC, \url{https://www.cosmos.esa.int/web/gaia/dpac/consortium}). Funding for the DPAC has been provided by national institutions, in particular the institutions participating in the {\it Gaia} Multilateral Agreement.

Funding for RAVE has been provided by: the Australian Astronomical Observatory; the Leibniz-Institut fuer Astrophysik Potsdam (AIP); the Australian National University; the Australian Research Council; the French National Research Agency; the German Research Foundation (SPP 1177 and SFB 881); the European Research Council (ERC-StG 240271 Galactica); the Istituto Nazionale di Astrofisica at Padova; The Johns Hopkins University; the National Science Foundation of the USA (AST-0908326); the W. M. Keck foundation; the Macquarie University; the Netherlands Research School for Astronomy; the Natural Sciences and Engineering Research Council of Canada; the Slovenian Research Agency; the Swiss National Science Foundation; the Science \& Technology Facilities Council of the UK; Opticon; Strasbourg Observatory; and the Universities of Groningen, Heidelberg and Sydney. The RAVE web site is at https://www.rave-survey.org.

This publication makes use of data products from the Two Micron All Sky Survey, which is a joint project of the University of Massachusetts and the Infrared Processing and Analysis Center/California Institute of Technology, funded by the National Aeronautics and Space Administration and the National Science Foundation.

\appendix

\section{Selection Function}\label{app:sel_fn}

The selection function of our data set is non-trivial. The Gaia selection is biased by the sky coverage offered by its orbital position and angle, this manifests as large stripes on the sky where there are too few observations to reliably report data. LAMOST sky coverage is focused in the direction of the anti-center, and certain fields, such as the Kepler field \citep{dec2015} enjoy even more dense coverage; adding to the confusion is the fact that PIs associated with the project can upload personal target lists which are then factored into the survey. RAVE is more uniform in its sky coverage, and is claimed to be kinematically and chemically unbiased from the expectations \citep{woj2017}, but it still has just one hemisphere of coverage and magnitude limits differing from LAMOST and Gaia.

To correct for this complex convolution of selection functions, we must define the goal of our correction: we are interested in a sample which represents the underlying population. Since we are investigating age, metallicity, and velocities, we cannot weight by these factors; instead we will weight by sky coverage (to account for, for example, the heavily sampled Kepler field overpowering the sparsely sampled central regions), and color-magnitude space (to account for, e.g., stellar types which are selected against by parallax requirements, like giants, or for possible color biases, like possible BHB searches being factored into the LAMOST survey). To calculate our weightings in these spaces, we will compare to synthetic data produced by Galaxia\footnote{http://galaxia.sourceforge.net/Galaxia3pub.html} \citep{sha2011} in accordance with the Besan\c{c}on Galaxy Model \citep{rob2003}.

To create an acceptable Galaxia model, we use the default parameter file, and change the r$_{max}$ value (the maximum observational radius of the model) to 0.55 kpc, and the apparent magnitude limits to 5, and 15. Note that the model is extinction-added to be consistent with the observational space using the internal 3-dimensional dust values of the model.

\begin{figure}
\includegraphics[width=\textwidth]{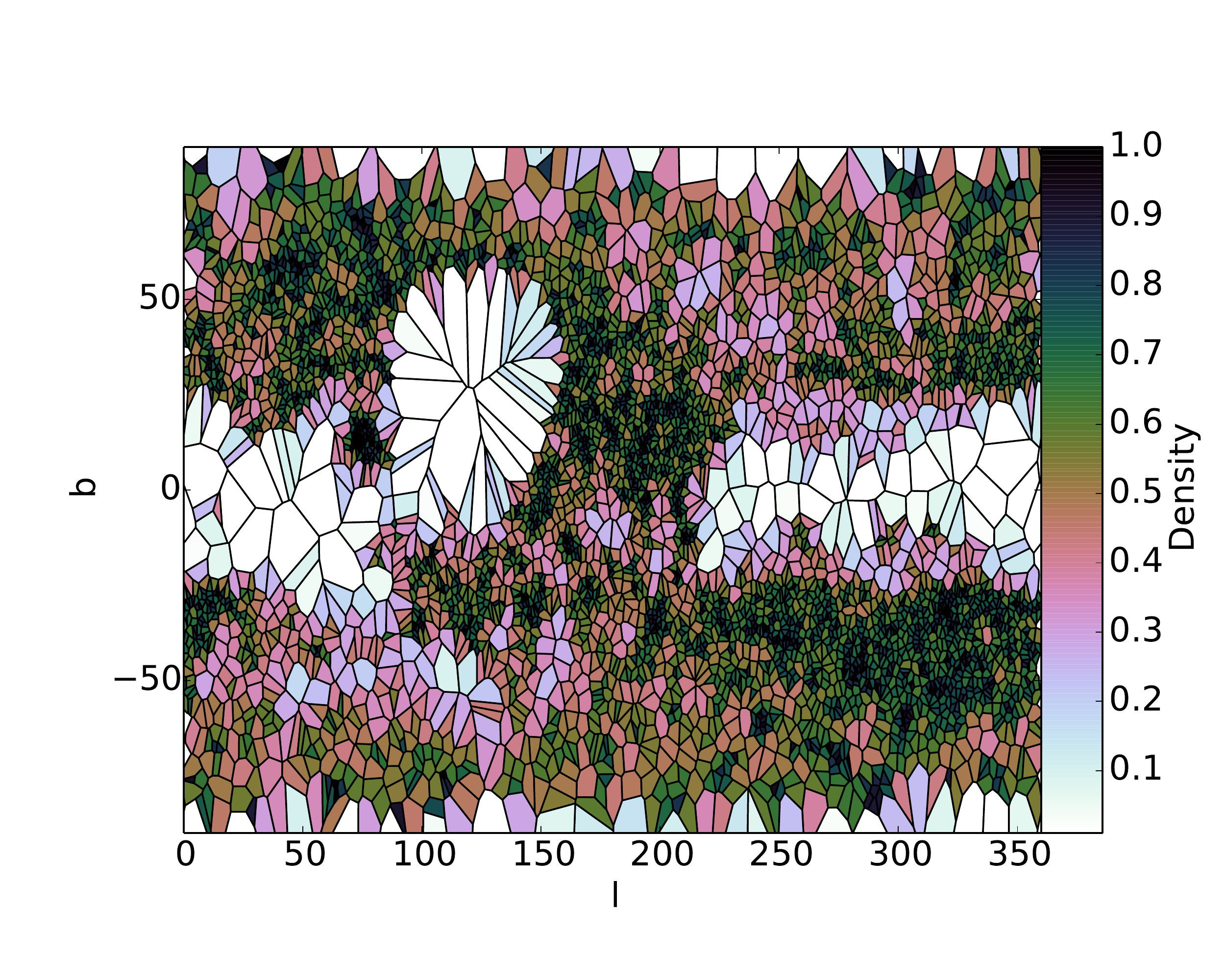}
\caption{The on-sky density of the observations, stars in lighter areas -- less well sampled areas -- are weighted more highly than stars in densely sampled areas. This is to prevent objects in, for example, the Kepler field, which is heavily sampled by LAMOST and may inadvertently sample some structure, from overpowering observations in the less crowded regions (e.g. the plane). Note that this image is merely illustrative and consists of just 5000 cells.}
\label{fig:vnoi}
\end{figure}

\begin{figure}
\includegraphics[width=\textwidth]{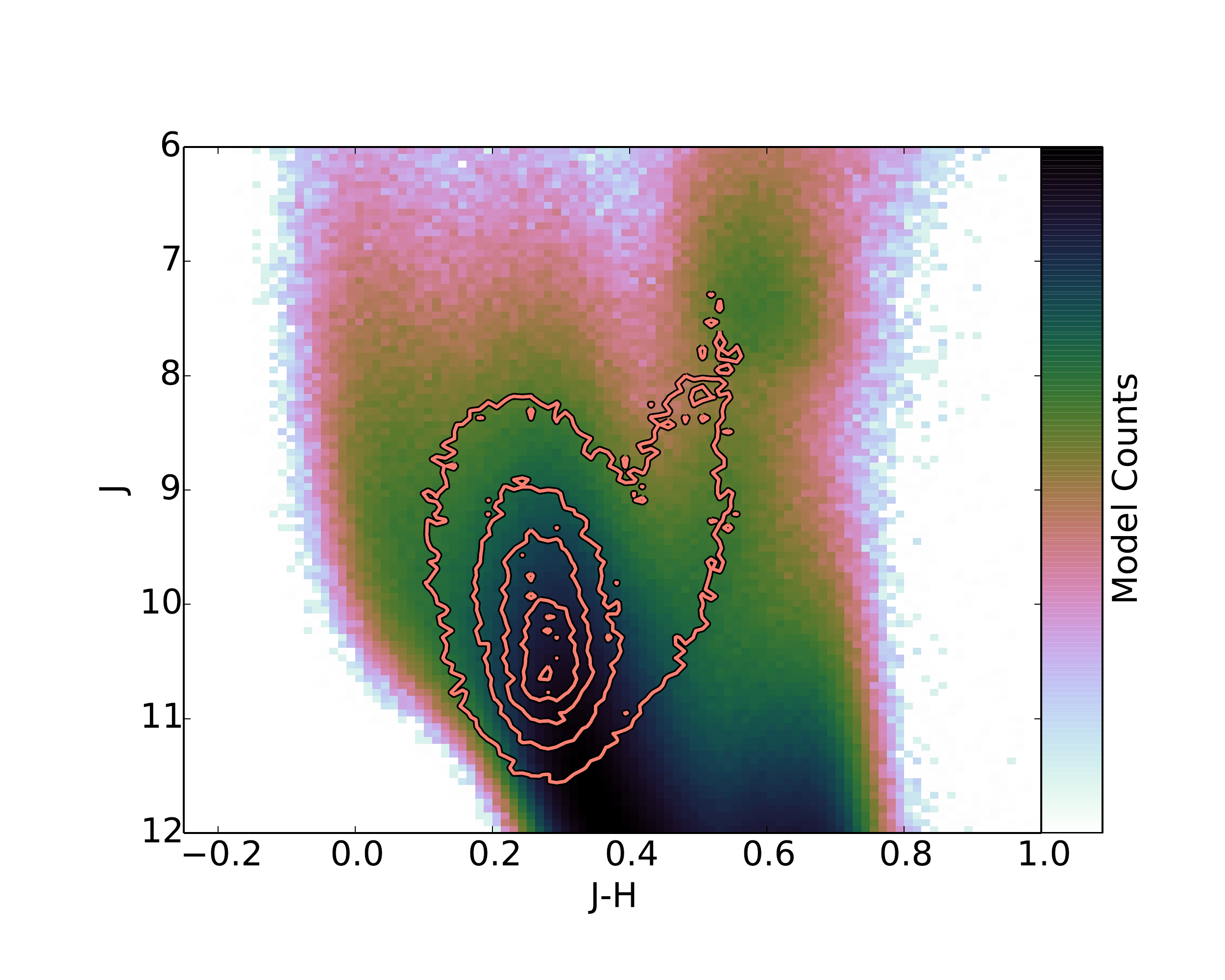}
\caption{A Hess diagram of our chosen Galaxia model overlain with the 5\%, 25\%. 50\%, 75\%, and 95\% density contours of our TGAS-LAMOST-RAVE sample. The Galaxia model is the default, with the apparent magnitude values chosen such that this diagram is complete in the range of the observations (see text). \\
}
\label{fig:galaxia}
\end{figure}

The weighting first accounts for on-sky density of observations by calculating a Voronoi cell for each star; a Voronoi cell is the area which is closer to the data point defining the cell than to any other data point (see Figure \ref{fig:vnoi}). So the inverse of the area of this cell is the density of observations at this individual observation; weighting by the cosine of the latitude accounts for the transformation of the Voronoi cell from cartesian to spherical space. Our observational coverage weight is:

\begin{equation}\label{eqn:weight_voronoi}
W_{Cell} = \frac{cos(b)}{ \rho_{Cell} }.
\end{equation}

Next we account for the expected on-sky density of the observation, normalized (over the model), using our simulated galaxy:

\begin{equation}\label{eqn:weight_galaxy}
W_{Sky} = \rho_{Model}(l, b).
\end{equation}

And, finally, we correct for the expected population at a specific ($l$, $b$) coordinate by collecting the 1000 nearest points from the model at that ($l$, $b$) and construct a 10x10x10 color-magnitude-distance grid. The relative density in the cell of the observation compared to those of the other cells is the weight of the observation.

\begin{equation}\label{eqn:weight_cmd}
W_{CMD} = \rho_{Model}(J, J-H, l, b, d).
\end{equation}

For the color-magnitude comparison, we use Two Micron All Sky Survey \citep{skr2006} $J$ magnitude and $J-H$ color (Figure \ref{fig:galaxia} compares the color-magnitude diagrams of our data and our model). Errors as a function of magnitude are estimated for $J$, $H$ and $d$ from the TGAS observational crossmatch, and these errors are then added in to the Galaxia mock observations. The weights are then normalized to sum to one before being combined into a total weighting for each star:

\begin{equation}\label{eqn:weight_final}
W = W_{Cell} \cdot W_{Sky} \cdot W_{CMD}.
\end{equation}

This total weighting is then normalized such that the weights sum to one. This correction does not create an exact duplication of the Galaxia model in any space, as some stars are removed owing to extremely high (or low) weightings, while other stars have measurements that are inconsistent with the model and preclude them from being weighted.

\section{Putting the Data on the Same Scale}\label{app:scale}

\subsubsection{Metallicity}
In the data description, we have referred to the iron abundance [Fe/H] for the LAMOST data, and to the metallicity [M/H] for the RAVE data. The LAMOST pipeline only calculates [Fe/H] and the RAVE pipeline only calculates [M/H] for most stars (and [Fe/H] for a smaller subset). These measures are not equivalent and, if not brought onto similar scales, will produce systematic differences in the estimated ages. To correct this we apply a very basic shift to the LAMOST data such that:

\begin{equation} \label{eqn:lamost_met}
   [M/H] = \left\{
     \begin{array}{ll}
       0.86[Fe/H] - 0.46 & : [Fe/H] < -0.7  \\
       1.22[Fe/H] - 0.26 & : [Fe/H] \geq -0.7 \\
     \end{array}
   \right.
\end{equation}

\begin{figure}
\includegraphics[width=\linewidth]{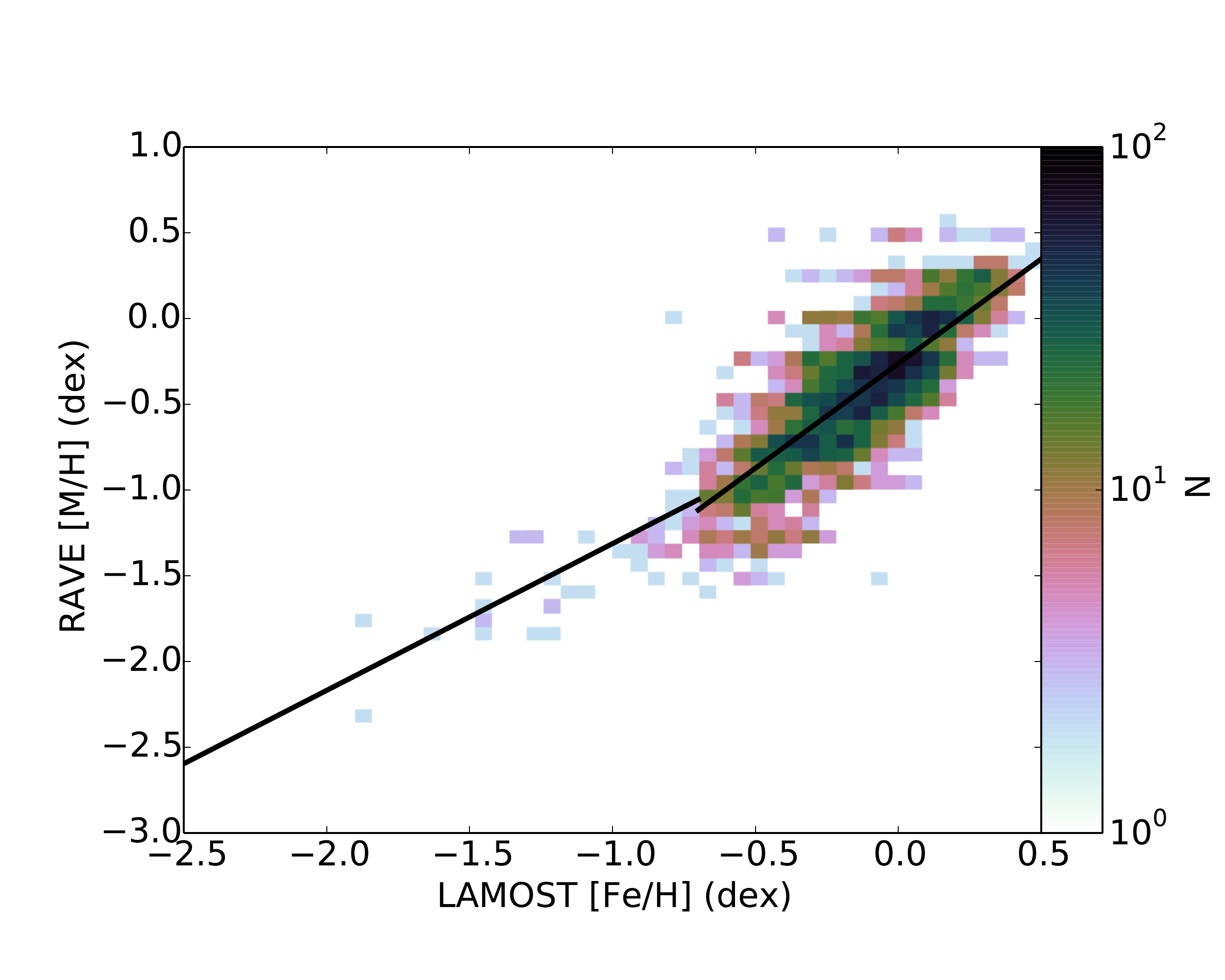}
\caption{LAMOST iron abundances compared to RAVE metallicity abundances in a set of $\sim$4100 objects which were observed by both RAVE and LAMOST. The black line is Equation \ref{eqn:lamost_met} which we use to bring LAMOST into a scale directly comparable to RAVE.}
\label{fig:feh_met}
\end{figure}

This shift is obtained by comparing overlapping observations in RAVE and LAMOST (we find about 4100 overlaps which match to within 1"). We then fit a piecewise function to the RAVE [M/H] parameter as a function of the LAMOST [Fe/H] parameter (see Figure \ref{fig:feh_met}). This shifts the two surveys into more equivalent scales, which will hopefully alleviate systematic biases between the two surveys when calculating ages. Rather than using [Fe/H] as a proxy for the full [M/H] content of a star when comparing to isochrones, we are shifting the LAMOST observations, based on the RAVE observations, to a new, hopefully more accurate estimate for the full [M/H] content of these stars. This also allows us to examine the relative age estimates for the RAVE and LAMOST samples together, rather than individually.

We have alpha abundance measurements for a subset of our RAVE data from the pipeline of \citet[see also \citealt{kun2017}]{boe2011}, but do not incorporate them into the age estimation. We have investigated the relative effects of alpha abundances on the age estimates, and find them to be similar to the effect of metallicity -- increasing the alpha abundances moves the star into a generally cooler, redder parameter space -- albeit, to a lesser extent. Since we do not have alpha-abundances for all of our objects, and since we wish to consider the LAMOST and RAVE data concurrently, we neglect the effects of alpha-abundances.

\subsubsection{Velocity}
It has been noted by \citet{tia2015} that LAMOST spectra have a systematic shift of about 5.7 km s$^{-1}$ when compared to radial velocities from APOGEE (a high-resolution, high signal to noise spectroscopic survey of red giants). It was also noted by \citet{sch2017} that the LAMOST radial velocities appeared to be underestimated by about 5 km s$^{-1}$. We note a similar offset in our LAMOST-RAVE overlapping data and apply the correction:

\begin{equation}\label{eqn:rv}
R.V._{LAMOST, C} = R.V._{LAMOST, P} + 5.7
\end{equation}

Where C and P stand for corrected value and pipeline value, respectively. As noted earlier, we also replace the LAMOST pipeline velocity errors with a uniform error of 7.2 km s$^{-1}$, which is the standard error found in \citet{sch2017}.

\end{document}